\documentclass[prb,floatfix,twocolumn,superscriptaddress]{revtex4}

\usepackage{amsmath}
\usepackage{graphicx}
\usepackage{bm}
\usepackage{hyperref}

\newcommand{\ket}[1]{\left \vert#1 \right \rangle}
 \def\bk{{\bf k}} 

\def\bV{{\bf V}}  
\def\bA{{\bf A}}  
\def\bE{{\bf E}} \def\bEDC{{\bf E}_{\mathrm{dc}}}
\def\bF{{\bf F}}
\def\bG{{\bf F}}

\def\bkperp{{\bf k}_\perp}  \def\zhat{\hat{\mathbf{z}}}

\def\Heff{H_{\mathrm{eff}}}
\def\bzhat{\hat{\mathbf{z}}}

\def\bp{{\bf p}}
\def\EDC{E_{\mathrm{dc}}}
\def\Eopt{E_{\mathrm{opt}}}\def\bEopt{{\bf E}_{\mathrm{opt}}}

\begin{document}
\title{Theory of the two-photon Franz-Keldysh effect and electric-field-induced bichromatic coherent control}
\author{J. K. Wahlstrand}
\email{jared.wahlstrand@nist.gov}
\affiliation{Nanoscale Device Characterization Division, National Institute of Standards and Technology, Gaithersburg, MD 20899 USA}
\author{J. E. Sipe}
\affiliation{Department of Physics, University of Toronto, Toronto, Ontario, Canada M5S1A7}
\date{\today}

\begin{abstract}
The effect of a constant electric field on two-photon absorption in a semiconductor is calculated using an independent-particle theory.
The theoretical framework is an extension of a theory of the one-photon Franz-Keldysh effect [Wahlstrand and Sipe, Phys.~Rev.~B \textbf{82}, 075206 (2010)].
The theory includes the effect of the constant field, including field-induced coupling between closely spaced bands, and with it we calculate the optical absorption perturbatively.
Numerical calculations are performed using a 14-band $\bk \cdot\bp$ band structure model for GaAs.
For all nonzero tensor elements, field-enabled two-photon absorption (TPA) below the band gap and Franz-Keldysh oscillations in the TPA spectrum are predicted, with a generally larger effect in tensor elements with more components parallel to the constant electric field direction.
Some tensor elements that are zero in the absence of a field become nonzero in the presence of the constant electric field and depend on its sign.
Notably, these elements are linear in the electric field to lowest order, and may be substantial away from band structure critical points at room temperature and/or with a non-uniform field.
Electric-field-induced changes in the carrier injection rate due to interference between one- and two-photon absorption are also calculated.
The electric field enables this bichromatic coherent control process for polarization configurations where it is normally forbidden by crystal symmetry, and also modifies the spectrum of the process for configurations where it is allowed.
\end{abstract}

\maketitle

\section{Introduction}

The Franz-Keldysh effect (FKE) describes the change in the absorption spectrum of a semiconductor material caused by a constant (DC) and uniform electric field, which accelerates electrons in their bands.\cite{franz_notitle_1958,keldysh_notitle_1958,aspnes_electric-field_1966}
The FKE is the mechanism behind electroreflectance and photoreflectance spectroscopy\cite{shen_franz--keldysh_1995} and some types of electroabsorption modulators.\cite{liu_waveguide-integrated_2008}
The analogous field-induced change in \emph{nonlinear} optical properties has received some attention \cite{kolber_theory_1978,garcia_tunneling_2006,garcia_phonon-assisted_2006,xia_franzkeldysh_2009,xia_nonlinear_2010,wahlstrand_polarization_2011,cui_modulation_2008,wahlstrand_optical_2011,wahlstrand_electric_2011} but remains a relatively obscure topic.
With increasing development of semiconductor-based integrated photonics,\cite{chang_integrated_2022,dutt_nonlinear_2024} much of which relies on nonlinear optics, it seems to be an opportune time to revisit the nonlinear optical aspect of the FKE because it could become important for engineering and optimizing devices.
For example, an optical switch that relies on the nonlinear interaction of multiple optical pulses could become electrically programmable using the nonlinear FKE or the related field-induced change in the optical Kerr coefficient $n_2$.
Even short of this novel potential application, the nonlinear optical FKE could also affect device performance in more subtle ways in integrated optical devices that utilize DC fields.
Bichromatic current injection -- arising from interference between one- and two-photon absorption of optical fields with frequencies $\omega$ and $2\omega$, where $2\hbar\omega$ is above the band gap\cite{atanasov_coherent_1996} -- has been used to measure and stabilize the carrier-envelope offset frequency of a frequency comb,\cite{fortier_carrier-envelope_2004,roos_solid-state_2005} and the closely related FKE-induced bichromatic control of carrier injection\cite{wahlstrand_optical_2011} might also be applied to modern integrated optics-based frequency combs.
A better fundamental understanding of the multiphoton FKE is essential to exploiting it for applications and accounting for its effects in modern devices.

In a previous paper we sketched out a theory of the two-photon FKE and calculated two-photon absorption (TPA) spectra using a two-parabolic-bands model, for which analytical solutions can be found, and an eight-band $\bk \cdot \bp$ model for GaAs (an archetypal direct band gap semiconductor).\cite{wahlstrand_polarization_2011}
Calculations using both the two-parabolic-bands model and the 8-band $\bk \cdot \bp$ model predicted that the modulation of the two-photon absorption coefficient depends strongly on the polarization of the optical field with respect to the applied DC field.\cite{garcia_tunneling_2006}
This theory was also used to calculate DC field-enabled bichromatic interference control and compared with experimental results.\cite{wahlstrand_optical_2011,wahlstrand_electric_2011}
Unlike the two-band model, which employs only a single matrix element, $\bk \cdot \bp$ models can lead to off-diagonal elements of the $\chi^{(3)}$ tensor, and thus to the prediction of nonvanishing two-photon absorption as a function of light polarization with respect to the crystal axes.
With enough parameters, these models can reflect the full symmetry of the crystal, including a lack of inversion symmetry, which leads to nonzero even-order nonlinear susceptibilities.

Previous publications that included calculations\cite{wahlstrand_polarization_2011,wahlstrand_optical_2011,wahlstrand_electric_2011} did not describe the complete theory.
Here we present the complete theory in detail, along with results of numerical calculations using a 14-band $\bk \cdot \bp$ model that, unlike the 8-band model, exhibits the lack of inversion symmetry of the crystal.
The theory presented here builds off our previously described theory of the one-photon FKE.\cite{wahlstrand_independent-particle_2010}
In realistic models with multiple bands, the theory of the Franz-Keldysh effect is complicated by coupling between bands near $\bk$ points and lines where bands are degenerate.\cite{hader_k.p_1997,wahlstrand_independent-particle_2010}
The DC field causes transitions between these bands as the wavevector of carriers passes near degeneracy points.
We handle this in the same way as was done previously\cite{wahlstrand_independent-particle_2010} and go to higher order in the optical perturbation to calculate two-photon absorption and bichromatic coherent control.
Both two-photon absorption and bichromatic interference are predicted to depend on the light polarization and the direction of the electric field with respect to the crystal axes.

\section{Theoretical Framework}
Our previous paper described a theoretical framework for calculating optical transitions perturbatively while accounting for DC field effects nonperturbatively.\cite{wahlstrand_independent-particle_2010}
The handling of the DC field is identical here, so we shall briefly summarize that, and then describe the extensions required to calculate two-photon absorption and quantum interference control.
Detailed derivations are provided in a supplemental document. 

\subsection{Interaction Hamiltonian}

We consider a crystal in the presence of a strong DC electric field $\bEDC$.
Assuming a uniform DC field, the electron wavefunctions within the crystal experience an acceleration.
As described earlier,\cite{wahlstrand_independent-particle_2010} we calculate using basis states with this acceleration ``built in.''
The component of the carrier wavevector parallel to the DC field is defined as $k_\parallel$ and is connected with time $t$ via $k_\parallel = \varepsilon t$, where $\varepsilon \equiv e \EDC/\hbar$ is a normalized electric field parameter.

To calculate optical absorption from an optical electric field $\bE_{\mathrm{opt}}$ we use an interaction picture, where a time-dependent ket evolves according to
\begin{equation}
  i\hbar \frac{d \left\vert \Psi(t) \right \rangle}{dt} = H_{\mathrm{eff}} \left\vert \Psi(t) \right \rangle,
\end{equation}
where the interaction Hamiltonian is
\begin{equation}
  H_{\mathrm{eff}}(t) = -\frac{1}{c} \bA_{\mathrm{opt}}(t) \cdot \mathcal{\tilde{J}}(t),
\end{equation}
with
\begin{equation*}
 \bA_{\mathrm{opt}}(t) = \int_{-\infty}^{\infty} \frac{d\omega}{2\pi} \bA_{\mathrm{opt}}(\omega) e^{-i\omega t},
\end{equation*}
where $\bA_{\mathrm{opt}}(\omega) = -ic\bE_{\mathrm{opt}}(\omega)/\omega$, and it is convenient to write
\begin{equation}
\bE_{\mathrm{opt}}(t) = \int_0^{\infty} \frac{d\omega}{2\pi} \bE_{\mathrm{opt}}(\omega) e^{-i\omega t} + c.c.,
\label{Eopt}
\end{equation}
and
\begin{equation}
\mathcal{\tilde{J}}(t) = e \sum_{n_1,n_2,\bk} b^\dagger_{n_2 \bk} b_{n_1 \bk} \tilde{\mathbf{V}}_{n_2 n_1}(\bk;t),
\end{equation}
with $b_{n \bk}$ and $b^\dagger_{n \bk}$ the annihilation and creation operators for electrons, and
\begin{equation}
   \tilde{\mathbf{V}}_{nq}(\bk;t) = \sum_{m,p} L^*_{mn}(\bk;t) \mathbf{V}_{mp}(\bk;t) L_{pq}(\bk;t).
\end{equation}
Here the velocity matrix elements $\mathbf{V}_{mp}(\bk;t)$ describe the usual coupling between the conduction and valence bands.
The evolution matrix $\mathrm{L}(\bk;t)$ includes the coupling between bands due to the DC field, which becomes particularly large at wavevectors where bands are nearly degenerate.\cite{wahlstrand_independent-particle_2010}
The matrix elements $\tilde{\mathbf{V}}_{nq}(\bk;t)$ are effective matrix elements that include this coupling.

\subsection{Perturbation calculation to second order}
To calculate one-photon absorption, we previously found\cite{wahlstrand_independent-particle_2010} $\ket{\Psi^{(1)}}$ in the iterative solution of the Schr\"{o}dinger equation
\begin{math}
\left\vert \Psi (t)\right\rangle =\left\vert \Psi ^{H}\right\rangle
+\left\vert \Psi ^{(1)}(t)\right\rangle + \ket{\Psi^{(2)} (t)} + \cdots ,
\end{math}
taking $\ket{\Psi^{(1)}}$ to denote the value of $\ket{\Psi^{(1)}(t)}$ at times after the optical pulse
has passed, with $\left\vert \Psi ^{H}\right\rangle$ as the initial condition.
Here we follow the same procedure but to second order,
\begin{equation}
\left\vert \Psi ^{(2)}\right\rangle =\frac{1}{(i\hbar)^2 }\int_{-\infty}^{\infty} \Heff (t^\prime) \int_{-\infty
}^{t^\prime}H_{\mathrm{eff}}(t^{\prime \prime})\left\vert \Psi ^{H}\right\rangle
dt^{\prime\prime} dt^{\prime}.  \label{psi2}
\end{equation}%
It is shown in the supplemental document that
\begin{widetext}
\begin{equation}
\left\vert \Psi^{(2)}\right\rangle  \label{psi2DC} 
= \sum_{c,v,\mathbf{k}_{\perp },k_{\parallel }}\iint d\omega _{a}d\omega
_{d}\theta _{cv\mathbf{k}_{\perp }}^{ij}(\omega _{a},\omega
_{d})\Eopt^{i} \left(\omega _{a}-\frac{1}{2}\omega _{d} \right)\Eopt^{j}\left( \omega _{a}+\frac{1%
}{2}\omega _{d}\right) e^{2i\omega _{a}k_{\parallel }/\varepsilon
}\left\vert \overline{cv(\mathbf{k}_{\perp }k_{\parallel })}\right\rangle ,
\end{equation}
where $\left\vert \overline{cv(\mathbf{k}_{\perp }k_{\parallel })}\right\rangle$ is a state that corresponds to an electron removed from band $v$ and placed in band $c$ at $\bk = \bkperp + \zhat k_\parallel$,\cite{wahlstrand_independent-particle_2010} and
\begin{multline}
\theta _{cv\mathbf{k}_{\perp }}^{ij}(\omega _{a},\omega _{d})
=\frac{ie^{2}}{4\pi^3 \hbar ^{2}(4\omega _{a}^{2}-\omega _{d}^{2})} \sum_{n}\int \frac{d\omega_{m}}{\omega _{d}-\omega_{m}} \left[\;F_{cn}^{i}(\mathbf{k}_{\perp };-\omega _{a}+\frac{1}{2}\omega_{m})F_{nv}^{j}(\mathbf{k}_{\perp };-\omega _{a}-\frac{1}{2}\omega_{m}) \right. \\ \left. -F_{cn}^{j}(\mathbf{k}_{\perp };-\omega _{a}-\frac{1}{2}\omega_{m}) F_{nv}^{i}(\mathbf{k}_{\perp };-\omega _{a}+\frac{1}{2}\omega_{m})\right] \label{thetatwo},
\end{multline}
in which
\begin{eqnarray}
\bG_{mn}(\bk_\perp; -\omega) &=& \int_{-\infty}^{\infty} dt \bG_{mn} (\bkperp; t) e^{-i\omega t}, \\
\label{gamma}
\mathbf{F}_{cv}(\mathbf{k}_{\perp };t)&=&
\sum_{c^{\prime },v^{\prime }}m_{c^{\prime }c}^{\ast }(\mathbf{k}_{\perp };t)\mathbf{V}_{c^{\prime }v^{\prime }}(\mathbf{k}_{\perp };t)m_{v^{\prime }v}(\mathbf{k}_{\perp };t),  \label{Fname}
\end{eqnarray}
where $\mathrm{m}(\bkperp; t)$ is closely related to the evolution matrix $\mathrm{L}(\bk; t)$.\cite{wahlstrand_independent-particle_2010}

\subsection{Two-photon absorption in the presence of a DC field}

For two-photon absorption, we calculate the number of carriers injected
\begin{align}
\Delta N =& \left\langle \Psi^{(2)} \vert \Psi^{(2)} \right \rangle \notag \\
=&\sum_{c,v,\mathbf{k}_{\perp },k_{\parallel }}\int d\omega _{a}d\omega
_{d}d\omega _{a}^{\prime }d\omega _{d}^{\prime } \theta _{cv\mathbf{k}%
_{\perp }}^{ij}(\omega _{a},\omega _{d}) \left[ \theta _{cv\mathbf{k}%
_{\perp }}^{lm}(\omega _{a}^{\prime },\omega _{d}^{\prime })\right]^{\ast } \notag
\\ &\times E^{i} \left(\omega _{a}-\frac{1}{2}\omega _{d} \right)E^{j}\left( \omega _{a}+%
\frac{1}{2}\omega _{d}\right) \left[ E^{l} \left(\omega _{a}^{\prime }-\frac{1}{2}%
\omega _{d}^{\prime } \right)E^{m}\left( \omega _{a}^{\prime }+\frac{1}{2}\omega
_{d}^{\prime }\right) \right] ^{\ast }e^{2i(\omega _{a}-\omega _{a}^{\prime
})k_{\parallel }/\varepsilon },
\end{align}
where we have used the fact that the states $\overline{\left\vert cv(\mathbf{k}_{\perp }k_{\parallel })\right\rangle }$ are orthonormal, and we have dropped the ``opt'' subscript on the optical electric field $\bE_{\mathrm{opt}}$.
Converting the sum over $\mathbf{k}_{\perp }$ and $k_{\parallel }$ to integrals, we calculate the number of carriers injected per unit volume,
\begin{eqnarray*}
\Delta n &=&\sum_{c,v}\int \frac{d\mathbf{k}_{\perp }}{4\pi^2}\int d\omega
_a d\omega_d d\omega_a^{\prime }d\omega_d^{\prime } \theta
_{cv\mathbf{k}_{\perp }}^{ij}(\omega _{a},\omega _{d}) \left[ \theta
_{cv\mathbf{k}_{\perp }}^{lm}(\omega _{a}^{\prime },\omega _{d}^{\prime
})\right] ^{\ast } \\
&&\times E^{i} \left(\omega _{a}-\frac{1}{2}\omega _{d} \right)E^{j}\left( \omega _{a}+
\frac{1}{2}\omega _{d}\right) \left[ E^{l} \left(\omega _{a}^{\prime }-\frac{1}{2}
\omega _{d}^{\prime } \right)E^{m}\left( \omega _{a}^{\prime }+\frac{1}{2}\omega
_{d}^{\prime }\right) \right]^\ast \int \frac{dk_{\parallel }}{2\pi }
e^{2i(\omega _{a}-\omega _{a}^{\prime })k_{\parallel }/\varepsilon }.
\end{eqnarray*}
The $k_{\parallel }$ integration gives a delta function for $\omega
_{a}$ and $\omega _{a}^{\prime }$, yielding
\begin{multline}
\Delta n =\int_{0}^{\infty }d\omega _{a}\int d\omega _{d}d\omega _{d}^{\prime} \left( \sum_{c,v}\frac{\varepsilon }{2}\int \frac{d\mathbf{k}_{\perp }}{4\pi^2}\theta _{cv\mathbf{k}_{\perp }}^{ij}(\omega _{a},\omega_{d}) \left[ \theta _{cv\mathbf{k}_{\perp }}^{lm}(\omega _{a},\omega
_{d}^{\prime })\right] ^{\ast }\right)  \\ \times E^{i} \left( \omega _a-\frac{1}{2}\omega_d \right)E^{j}\left( \omega _{a}+\frac{1}{2}\omega _{d}\right) \left[ E^{l} \left(\omega _{a}-\frac{1}{2}\omega
_{d}^{\prime } \right)E^{m}\left( \omega _{a}+\frac{1}{2}\omega _{d}^{\prime
}\right) \right]^{\ast },  \label{2pDC} 
\end{multline}
where we have used Eq.~(\ref{Eopt}).

\subsection{Bichromatic coherent control}
When frequency components $\omega$ and $2\omega$ are present, there are interference terms between one- and two-photon absorption.
In all we have
\begin{eqnarray*}
\Delta N &=&\left( \left\langle \Psi^{(1)}\right\vert +\left\langle
\Psi^{(2)}\right\vert \right) \left( \left\vert \Psi^{(1)}\right\rangle +\left\vert \Psi^{(2)}\right\rangle \right)
\\
&=&\Delta N_{(1)}+\Delta N_{(2)}+\Delta N_{(I)},
\end{eqnarray*}
where
\begin{math}
\Delta N_{(1)} =\left\langle \Psi^{(1)}|\Psi^{(1)}\right\rangle,
\end{math}\cite{wahlstrand_independent-particle_2010}
\begin{math}
\Delta N_{(2)} =\left\langle \Psi^{(2)}|\Psi^{(2)}\right\rangle
\end{math}
was calculated in the previous section, and the interference term
\[
\Delta N_{(I)}=\left\langle \Psi^{(2)}|\Psi^{(1)}\right\rangle
+c.c. 
\]

We combine Eq.~(\ref{psi2DC}) with the one-photon ket\cite{wahlstrand_independent-particle_2010}
\begin{equation}
\left\vert\Psi^{(1)} \right\rangle = \sum_{c,v,\bkperp,k_\parallel} \int d\omega \theta^i_{cv\bkperp}(\omega) E^i(\omega) e^{i\omega k_\parallel/\varepsilon} \left\vert \overline{cv(\mathbf{k}_{\perp }k_{\parallel })}\right\rangle,
\end{equation}
where
\begin{equation}
  \theta_{cv\bkperp}^i (\omega) = \frac{e}{2\pi\hbar} \frac{\bF_{cv}^i (\bkperp; -\omega)}{\omega},
  \label{thetaone}
\end{equation}
to give 
\begin{eqnarray*}
\Delta N_{(I)} 
&=&\sum_{c,v,\mathbf{k}_{\perp },k_{\parallel }}\int d\omega _{a}d\omega
_{d}d\omega \left[ \theta _{cv\mathbf{k}_{\perp }}^{ij}(\omega _{a},\omega
_{d})\right] ^{\ast }\theta _{cv\mathbf{k}_{\perp }}^{l}(\omega ) \\
&&\times \left[ E^{i} \left(\omega _{a}-\frac{1}{2}\omega _{d}\right)E^{j}\left( \omega
_{a}+\frac{1}{2}\omega _{d}\right) \right] ^{\ast }E^{l}(\omega )e^{i(\omega
-2\omega _{a})k_{\parallel }/\varepsilon } +c.c.
\end{eqnarray*}%
Converting the sums over $\mathbf{k}_{\perp }$ and $k_{\parallel }$ to
integrals, we have 
\begin{eqnarray*}
\Delta n_{(I)}
&=&\int d\omega _{a}d\omega _{d}d\omega \sum_{c,v}\int \frac{d\mathbf{k}%
_{\perp }}{4\pi ^{2}}\frac{dk_{\parallel }}{2\pi }\left[ \theta _{cv\mathbf{k%
}_{\perp }}^{ij}(\omega _{a},\omega _{d})\right] ^{\ast }\theta _{cv\mathbf{k%
}_{\perp }}^{l}(\omega ) \\
&&\times \left[ E^{i} \left(\omega _{a}-\frac{1}{2}\omega _{d} \right)E^{j}\left( \omega
_{a}+\frac{1}{2}\omega _{d}\right) \right] ^{\ast }E^{l}(\omega )e^{i(\omega
-2\omega _{a})k_{\parallel }/\varepsilon } +c.c.
\end{eqnarray*}%
The integral over $k_{\parallel }$ can now be done to yield
\begin{multline}
\Delta n_{(I)} 
=\varepsilon \int d\omega _{a}d\omega _{d}d\omega \delta (\omega -2\omega
_{a})  \sum_{c,v}\int \frac{d\mathbf{k}_{\perp }}{4\pi ^{2}}\left[ \theta
_{cv\mathbf{k}_{\perp }}^{ij}(\omega _{a},\omega _{d})\right] ^{\ast }\theta
_{cv\mathbf{k}_{\perp }}^{l}(\omega ) \\ \times \left[ E^{i} \left(\omega _{a}-\frac{1}{2}\omega _{d} \right)E^{j}\left( \omega
_{a}+\frac{1}{2}\omega _{d}\right) \right] ^{\ast }E^{l}(\omega ) +c.c., \nonumber
\end{multline}
or 
\begin{multline}
\Delta n_{(I)}  \label{interDC} 
=\varepsilon \int d\omega _{a}d\omega _{d}\sum_{c,v}\int \frac{d\mathbf{k}%
_{\perp }}{4\pi ^{2}}\left[ \theta _{cv\mathbf{k}_{\perp }}^{ij}(\omega
_{a},\omega _{d})\right] ^{\ast }\theta _{cv\mathbf{k}_{\perp }}^{l}(2\omega
_{a}) \\ \times \left[ E^{i} \left(\omega _{a}-\frac{1}{2}\omega _{d} \right)E^{j}\left( \omega
_{a}+\frac{1}{2}\omega _{d}\right) \right] ^{\ast }E^{l}(2\omega _{a}) +c.c.
\end{multline}

\subsection{Fermi's Golden Rule limit}
The expressions given so far hold for arbitrary optical pulses.
In the case of two-photon transitions the pulse shape can strongly affect the amount of carrier injection,\cite{meshulach_coherent_1998} and the two-photon injection rate can be very different for highly nondegenerate excitation.\cite{fishman_sensitive_2011}
Nevertheless, since it is often the nonlinearity in the response to a single incident beam that is measured, the continuous wave (CW) limit is of practical interest.

Consider a continuous wave field of the form $\bE = \bE_o e^{-i\omega_o t} + \bE_o^* e^{i\omega_o t}$.
For two-photon absorption, we can use, as shown in the supplemental document,
\begin{equation}
E^{i}(\omega +\frac{1}{2}\omega _{d})E^{j}(\omega -\frac{1}{2}\omega
_{d})\left( E^{l}(\omega +\frac{1}{2}\omega _{d}^{\prime })E^{m}(\omega -%
\frac{1}{2}\omega _{d}^{\prime })\right) ^{\ast }
\rightarrow 16\pi ^{3}T\;E_{o}^{i}E_{o}^{j}\left(
E_{o}^{l}E_{o}^{m}\right) ^{\ast }\delta (\omega _{d})\delta (\omega
_{d}^{\prime })\delta (\omega -\omega _{o}).  \label{2photondelta}
\end{equation}
Inserting Eq.~(\ref{2photondelta}) into Eq.~(\ref{2pDC}), we can define a polarization-dependent carrier injection tensor for two-photon absorption
\begin{equation}
\eta^{ijlm}_2 (\omega_a; \bEDC) = 16\pi^3 \sum_{cv}\frac{\varepsilon }{2}\int \frac{d\mathbf{k}_{\perp }}{4\pi^2}\left[ \theta _{cv\mathbf{k}_{\perp }}^{ij}(\omega _{a},0)\right] \left[ \theta _{cv\mathbf{k}_{\perp }}^{lm}(\omega _{a},0)\right] ^{\ast },
\label{eta}
\end{equation}
(the right hand side depends implicitly on $\bEDC$ through $\varepsilon$ and $\theta$) such that
\begin{equation}
\frac{dn_{(2)}}{dt}= \eta^{ijlm}_2 (\omega_o; \bEDC) E^i_o E^j_o (E^l_o E^m_o)^*.
\end{equation}

It is shown in the supplemental document that with no electric field, the carrier injection tensor is
\begin{equation}
\eta^{ijlm}_2 (\omega_a; \bm{0}) =  16\pi^3 \sum_{cv} \int \frac{d\bk}{8\pi^3}\frac{1}{2} \gamma_{cv\bk}^{ij}(0)  \left( \gamma_{cv\bk}^{ij}(0) \right)^* \delta(2\omega_a - \omega_{cv}(\bk)),
\end{equation}
where
\begin{equation}
\gamma _{cv\mathbf{k}}^{ij}(\omega _{d}) 
=\frac{2ie^{2}}{\pi \hbar ^{2}(\omega _{cv}^{2}(\mathbf{k})-\omega
_{d}^{2})} \sum_{n}\left( \frac{v_{cn}^{i}(\mathbf{k})v_{nv}^{j}(\mathbf{k})}{%
\left( \omega _{cn}(\mathbf{k})+\omega _{vn}(\mathbf{k})+\omega _{d}\right) }%
\right.  + \left. \frac{v_{cn}^{j}(\mathbf{k})v_{nv}^{i}(\mathbf{k})}{\left( \omega _{cn}(%
\mathbf{k})+\omega _{vn}(\mathbf{k})-\omega _{d}\right) }\right).
\label{gamma_two}
\end{equation}
This agrees with the expression derived for a previous $\bk \cdot \bp$ calculation.\cite{rioux_optical_2012}

For bichromatic interference control in the continuous wave limit, we consider a field that can be described as consisting of joint pulses
involving a carrier at $\omega _{o}$ and a carrier at $2\omega _{o}$, writing
\begin{equation}
\bE(t) =\mathbf{E}_{F}e^{-i\omega _{o}t}+\mathbf{E}_{F}^{\ast
}e^{i\omega _{o}t}  
+\mathbf{E}_{S}e^{-2i\omega _{o}t}+\mathbf{E}_{S}^{\ast }e^{2i\omega _{o}t}.
\end{equation}
It is shown in the supplemental document that in the continuous wave limit,
\begin{equation}
E_{F}^{l}(\omega +\frac{1}{2}\omega _{d})E_{F}^{m}(\omega -\frac{1}{2}
\omega _{d})\left( E_{S}^{i}(2\omega )\right) ^{\ast }  \label{inter} \rightarrow 4\pi ^{2}TE_{F}^{l}E_{F}^{m}\left( E_{S}^{i}\right) ^{\ast
}\delta (\omega _{d})\delta (\omega -\omega _{o}).
\end{equation}
Finally, using Eq.~(\ref{inter}) in Eq.~(\ref{interDC}) and $\bE_F = \bE_\omega e^{i\phi_\omega}$ and $\bE_S = \bE_{2\omega} e^{2i\phi_{2\omega}}$ for $\bE_\omega$ and $\bE_{2\omega}$ real, we find
\begin{equation}
\frac{dn_{(I)}}{dt} 
= \eta^{ijl}_I(\omega_o; \bEDC) \left[ E_{F}^{i}E_{F}^{j} \right] ^{\ast }E_{S}^{l},
\end{equation}
where the carrier injection tensor for interference control is
\begin{equation}
\eta^{ijl}_I(\omega_a; \bEDC) = 4\pi ^{2}\varepsilon \sum_{cv}\int \frac{d\mathbf{k}_{\perp }}{4\pi ^{2}}%
\left[ \theta _{cv\mathbf{k}_{\perp }}^{ij}(\omega_a,0)\right] ^{\ast
} \theta _{cv\mathbf{k}_{\perp }}^{l}(2\omega_a)e^{-2i\phi_\omega+i\phi_{2\omega}} +c.c.
\end{equation}

It is shown in the supplemental document that with no DC electric field the carrier injection tensor is
\begin{equation}
\eta^{ijl}_I(\omega_a; \bm{0}) = 
4\pi ^{2}\left( \sum_{cv}\int \frac{d\mathbf{k}}{8\pi ^{3}}\left[ \gamma
_{cv\mathbf{k}}^{ij}(0)\right] ^{\ast }\gamma _{cv\mathbf{k}}^{l}\delta
(2\omega_a-\omega_{cv}(\mathbf{k}))\right) +c.c.,
\end{equation}
where $\gamma_{cv\mathbf{k}}^{ij}(\omega_d)$ is given by Eq.~(\ref{gamma_two}) and $\gamma^i_{cv\bk} = ev_{cv}^i/[\hbar \omega_{cv}(\bk)]$.

\end{widetext}

\section{Calculations}
For photon energies near $E_g/2$, two-photon absorption in GaAs and other direct zincblende semiconductors is largely due to transitions involving two bands, so models that account for only two bands can qualitatively capture the onset of absorption.\cite{catalano_interband_1988}
However, processes that involve a third band also contribute significantly to two-photon absorption,\cite{pidgeon_two-photon_1979} and the shape of bands also affects the absorption spectrum.
The $\bk \cdot \bp$ method\cite{lax_symmetry_1974} provides a band structure that accurately describes the relative band energies and the symmetry of the states as a function of $\bk$ within some range of energies near the fundamental band gap.
For two-photon calculations we previously used an 8-band model,\cite{wahlstrand_polarization_2011} which includes the six uppermost valence bands and the two lowermost conduction bands.
Here we use a 14-band model with six additional higher energy conduction bands that contribute nontrivially to the two-photon absorption, particularly for certain elements of $\eta^{ijlm}$.
The six additional upper conduction bands included in the 14-band model also modify the symmetry of the valence and lower conduction bands, introducing effects that only appear in the absence of a center of inversion symmetry.

Details of the 14-band $\bk \cdot \bp$ model implementation used here are given by Pfeffer and Zawadzki,\cite{pfeffer_five-level_1996} Bhat and Sipe,\cite{bhat_calculations_2006} and Wahlstrand and Sipe.\cite{wahlstrand_independent-particle_2010}
To summarize, $\bk\cdot\bp$ model parameters generally include band energies at the $\Gamma$ point, couplings between bands that determine effective masses near critical points, and optionally additional parameters that further refine the band dispersion.
In GaAs, energy parameters at $\bk = 0$ include the band gap $E_g = 1.519$~eV, the separation between the split-off valence band and the highest valence bands $\Delta_0 = 0.341$~eV, the gap between the highest valence bands and the upper conduction bands $E_0^\prime = 4.488$~eV, and the upper conduction band spin splitting $\Delta_0^\prime = 0.171$~eV.
The model includes three coupling parameters for GaAs: $P_0 = 10.3$~eV$\cdot\mathrm{\AA}$ is the coupling between the valence bands and the lowest conduction bands, $Q = 7.7$~eV$\cdot\mathrm{\AA}$ is the coupling between the valence bands and the upper conduction bands, and $P_0^\prime=3.0$~eV$\cdot\mathrm{\AA}$ is the coupling between the lower and higher conduction bands.
An additional spin splitting parameter $\Delta^- = -0.061$~eV couples the valence bands and upper conduction bands.
Remote band effects on the valence bands are included through modified Luttinger parameters $\gamma_1 = 7.797$, $\gamma_2 = 2.458$, and $\gamma_3 = 3.299$, which adjust the hole effective masses to match observed values.
Additional parameters further improve accuracy: $F = -1.055$ fixes the conduction band effective mass to the observed value and $C_k = -0.0034$~eV$\cdot \mathrm{\AA}$ adjusts the spin splitting.\cite{bhat_calculations_2006}

Each expression for the carrier injection tensor for non-zero $\bEDC$ consists of an integral over $\bkperp$.
The calculation is performed for one value of $\bkperp$ at a time.
First, time-dependent matrix elements $\bF_{mn}(\bkperp; t)$, which include transitions between nearly degenerate bands, are calculated as described in the one-photon FKE calculation.\cite{wahlstrand_independent-particle_2010}
As in our previous paper, remote band effects are included in these matrix elements for consistency.\cite{bhat_calculations_2006,enders_kp_1995}
Next, the quantities $\theta_{cv\bkperp}^{i}(\omega_a, 0)$ and $\theta_{cv\bkperp}^{ij}(\omega)$ are calculated using Eqs.~(\ref{thetaone}) and (\ref{thetatwo}), respectively.
The two-photon calculation is considerably more time consuming than the one-photon calculation because Eq.~(\ref{thetatwo}) requires integrating over $\omega_m$ and summing over all possible intermediate bands.
We calculated spectra over the photon energy range 0.7 to 1.0~eV, which covers the onset of absorption at half the fundamental gap $E_g/2$ up to half the split-off gap $E_{so}/2 = (E_g+\Delta_0)/2$.
This energy range corresponds to wavevectors $|\bk| \lesssim 0.1$~$\mathrm{\AA}^{-1}$, where the 14-band model is accurate.

\section{Results}
In this section we present results of the calculations.
Experimentally accessible quantities such as the electroabsorption spectrum, i.e.~the change in absorption spectrum due to the DC field, are calculated.
Note that as in our previous one-photon calculations,\cite{wahlstrand_independent-particle_2010} we neglect the Coulomb interaction between excited electrons and holes -- and thus effects due to bound electron-hole pairs, \emph{i.e.}, excitons -- as well as scattering and decoherence processes.
Duque-Gomez and Sipe calculated the one-photon absorption spectrum using a 14-band $\bk \cdot \bp$ model including the Coulomb interaction between the electron and hole.\cite{duque-gomez_franzkeldysh_2015}
They found a modification of the response at the band edge due to the Coulomb interaction, but the Franz-Keldysh oscillations had a magnitude and polarization dependence similar to that calculated without the Coulomb interaction.
We expect similar behavior for two-photon spectra.

\subsection{Two-photon absorption}
\begin{figure}
\includegraphics[width=8.5cm]{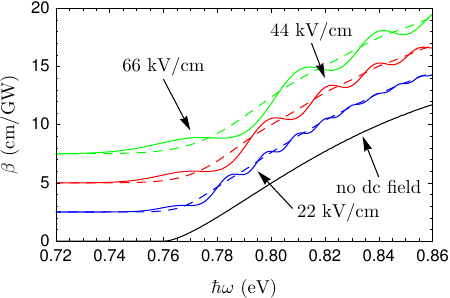}
\caption{Two-photon absorption spectra [Eq.~(\ref{beta})] in GaAs calculated using the 14-band $\bk \cdot \bp$ model without a field, and with a DC field pointing along $[001]$, for $\bEopt \parallel \bEDC$ (solid) and $\bEopt \parallel [100] \perp \bEDC$ (dashed).  The curves at each field strength are displaced vertically by 2.5~cm/GW.}
\label{kp_abs}
\end{figure}

In most experimental work, a two-photon absorption (TPA) coefficient $\beta(\omega)$ is reported.
This describes the change in absorption coefficient with irradiance, that is $\Delta \alpha = \beta I$, and $\beta$ is often given in units of cm/GW.
The TPA coefficient $\beta$ is proportional to the imaginary part of the third-order nonlinear susceptibility $\chi^{(3)}$, which is related to $\eta_2$ [Eq.~(\ref{eta})] via\cite{rioux_optical_2012}
\begin{equation}
  \mathrm{Im}\left[\chi^{(3)}_{ijlm}(\omega; -\omega, \omega, \omega) \right] = \frac{\hbar}{3} \eta^{ijlm}_2(\omega; \bEDC).
  \label{TPA_susceptibility}
\end{equation}
The TPA coefficient depends on the polarization state of the light with respect to the crystal axes.
For the DC field pointing along a crystal direction in zincblende crystals, it is given by\cite{sutherland_handbook_1996}
\begin{equation}
  \beta (\omega; \bEDC) = \frac{128\pi^5\hbar\omega}{n^2(\omega)c^2} \eta^{iiii}_2 (\omega; \bEDC),
  \label{beta}
\end{equation}
(no summation implied) for light linearly polarized in the $i$ direction, where $n(\omega)$ is the real refractive index.
Calculated TPA spectra are shown in Fig.~\ref{kp_abs}.
Absorption spectra for 22~kV/cm (blue), 44~kV/cm (red), and 66~kV/cm (green) applied DC fields along the [001] crystal direction are shown, along with the spectrum for zero DC field (black).
Solid lines show the absorption spectrum for light polarized parallel to the DC field, while dashed lines show the absorption spectrum for light polarized perpendicular to the DC field.
Our calculation with no DC field is essentially identical to a previous calculation using a 14-band model for GaAs.\cite{hutchings_theory_1994}
The DC field enables absorption below $E_g/2 = 0.76$~eV and produces Franz-Keldysh oscillations in the absorption spectrum above $E_g/2$, as previously found with a 2-parabolic-bands model and an 8-band $\bk \cdot \bp$ model.\cite{wahlstrand_polarization_2011}
The Franz-Keldysh oscillations are much more pronounced when the optical field is parallel to the DC field, which is also in agreement with our previous calculations.\cite{wahlstrand_polarization_2011}

\begin{figure}
\includegraphics[width=8.6cm]{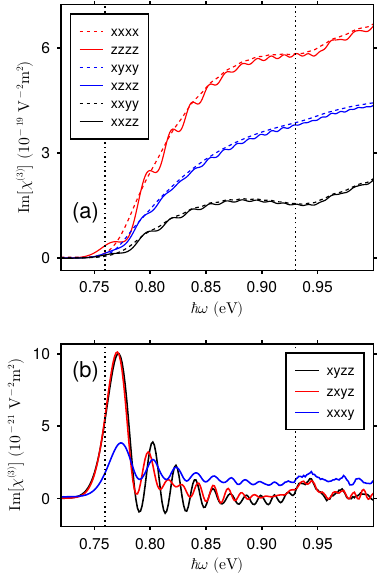}
\caption{Elements of $\mathrm{Im}[\chi^{(3)}]$ calculated using the 14-band $\bk \cdot \bp$ model for $\EDC = 44$~kV/cm along [001]. Dotted lines indicate $\hbar \omega = E_g/2$ and $\hbar\omega = E_{so}/2$. (a) Elements related to the even FKE. The solid lines are curves for the $zzzz$, $xzxz$, and $xxzz$ elements, which have at least one component pointing along the DC field direction. The dashed lines are for the $xxxx$, $xyxy$, and $xxyy$ elements, which have all components perpendicular to the DC field. (b) Elements related to the odd FKE: $xyzz$, $xxxy$, and $zxyz$.}
\label{offdiag}
\end{figure}

The rate of two-photon carrier injection depends on the polarization of the light with respect to the crystal directions.\cite{hutchings_theory_1994}
In the absence of a DC field, the zincblende crystal structure allows three independent nonzero elements in $\eta^{ijlm}_2$: $\eta^{xxxx}_2$, $\eta^{xyxy}_2$, and $\eta^{xxyy}_2$;\cite{hutchings_theory_1994} all permutations of these are the same since $x$, $y$, and $z$ are equivalent.
In the presence of a DC field along $\bzhat$, there are nine independent nonzero tensor elements: $\eta^{xxxx}_2$, $\eta^{zzzz}_2$, $\eta^{xxzz}_2$, $\eta^{xzxz}_2$, $\eta^{xxyy}_2$, $\eta^{xyxy}_2$, $\eta^{xyzz}_2$, $\eta^{xxxy}_2$, and $\eta^{zxyz}_2$, where $x$ and $y$ are now interchangeable.
Calculated tensor elements are plotted in Fig.~\ref{offdiag} as a function of photon energy for a DC field of 44~kV/cm along $\bzhat$.
To allow direct comparison with previous calculations of two-photon carrier injection in the absence of a DC field, we plot $\mathrm{Im}[\chi^{(3)}]$.\cite{rioux_optical_2012}
Vertical dotted lines in Fig.~\ref{offdiag} indicate locations of absorption edges $E_g/2$ and $E_{so}/2$.
The magnitude of the field-induced change in $\eta^{ijlm}_2$ depends strongly on whether one of the optical field components is along or perpendicular to the DC field.
The general trend appears to be that the more tensor elements are along the direction of the DC field, the stronger the effect.

For the DC field along $\bzhat$, six elements ($\eta^{xxxx}_2$, $\eta^{zzzz}_2$, $\eta^{xxzz}_2$, $\eta^{xzxz}_2$, $\eta^{xxyy}_2$, and $\eta^{xyxy}_2$) correspond to changes in nonzero tensor elements of $\eta_2$ in the absence of a DC field.
These tensor elements, which are shown in Fig.~\ref{offdiag}a, depend on the magnitude of the DC electric field but not its sign.
The electric field dependence of these tensor elements at a particular photon energy can thus be expressed as a series in the \emph{even} powers of the DC electric field, which is the usual situation for the Franz-Keldysh effect.
The other three nonzero tensor elements ($\eta^{xyzz}_2$, $\eta^{xxxy}_2$, and $\eta^{zxyz}_2$) are only non-zero in the presence of a DC electric field.
They are enabled by the breaking of crystal symmetry by the electric field.
These elements, which are shown in Fig.~\ref{offdiag}b, depend on the sign of the DC field and can be expressed as a series in \emph{odd} powers of the DC electric field.

As just discussed, for a DC field pointing along a crystal axis in GaAs certain elements have an even field dependence, while others have an odd field dependence.
If the field points along other directions, elements can have both even and odd components,
\begin{eqnarray}
\eta_{2,even}^{ijlm}(\omega; \bEDC) &=& \frac{\eta^{ijlm}_2(\omega; \bEDC) + \eta^{ijlm}_2(\omega; -\bEDC)}{2}, \nonumber \\
\eta_{2,odd}^{ijlm}(\omega; \bEDC) &=& \frac{\eta^{ijlm}_2(\omega; \bEDC) - \eta^{ijlm}_2(\omega; -\bEDC)}{2}. \nonumber
\end{eqnarray}
The field-induced change in absorption can be small, and it is useful to consider the \emph{electroabsorbance} spectrum, where the zero-field absorbance spectrum is subtracted off.
In experiments, the DC field is often modulated at a particular frequency and the change in absorption is detected using a lock-in amplifier.
In this scheme, the even FKE can be isolated by measuring at the second harmonic of the DC field modulation,\cite{wahlstrand_uniform-field_2010} while the odd FKE can be isolated by measuring at the fundamental. 
To isolate these two effects in the calculations, we calculate spectra using positive and negative DC fields $\bEDC$ and $-\bEDC$ and find even and odd electroabsorption spectra
\begin{equation}
  \Delta \eta_{2,even}^{ijlm}(\omega; \bEDC) = \eta_{2,even}^{ijlm}(\omega; \bEDC) - \eta_{2,even}^{ijlm}(\omega; \bm{0}) 
  \label{evenFKE}
\end{equation}
and the odd electroabsorption spectrum using
\begin{equation}
  \Delta \eta_{2,odd}^{ijlm}(\omega; \bEDC) = \eta_{2,odd}^{ijlm}(\omega; \bEDC) - \eta_{2,odd}^{ijlm}(\omega; \bm{0}) 
  \label{oddFKE}
\end{equation}
For consistency and to enable easy comparison with previous calculations, we plot the corresponding even and odd changes in $\mathrm{Im}\left[\chi^{(3)}\right]$ using Eq.~(\ref{TPA_susceptibility}).


\subsubsection{Even FKE}
\begin{figure}
\includegraphics[width=8.5cm]{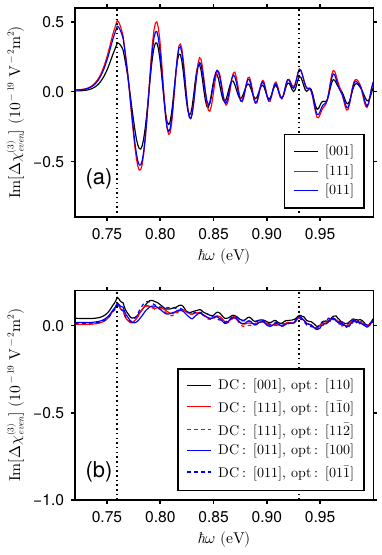}
\caption{Even two-photon electroabsorption spectra [Eq.~(\ref{evenFKE})] calculated using the 14-band $\bk \cdot \bp$ model in GaAs. All calculations used $\EDC = 44$~kV/cm. Dotted lines indicate $\hbar \omega = E_g/2$ and $\hbar\omega = E_{so}/2$. (a) Spectra for the optical field polarized parallel to the DC field direction. (b) Spectra for the optical field polarized perpendicular to the DC field.}
\label{electroabs_even}
\end{figure}

Even electroabsorption spectra calculated using Eq.~(\ref{evenFKE}) for a DC field of strength 44~kV/cm are shown in Fig.~\ref{electroabs_even}.
Figure \ref{electroabs_even}a shows spectra for the DC and optical fields parallel and along the [100], [110], and [111] crystal directions.
The Franz-Keldysh oscillation period is related to the electro-optic frequency $\Omega_{cv} = (\hbar \varepsilon^2/2\mu_{cv})^{1/3}$, where $\mu_{cv}$ is the reduced mass.\cite{aspnes_band_1974}
Because of selection rules, only the light-hole valence band contributes for parallel DC and optical fields,\cite{hader_k.p_1997,wahlstrand_independent-particle_2010} while for most other polarization configurations both heavy and light hole bands contribute, leading to beats in the Franz-Keldysh oscillations.
We also find that, as with the one-photon effect,\cite{wahlstrand_independent-particle_2010} the period of the oscillations depends slightly on the direction of the DC field because of the dependence of the effective mass on direction due to band warping.

The lineshapes for energies near $E_g/2$ and $E_{so}/2$ are generally consistent with the analytical expressions previously found for the two-band model assuming parabolic bands (PB).\cite{wahlstrand_polarization_2011}
The most significant difference is that the oscillations damp more quickly with increasing $\hbar \omega$ for the $\bk \cdot \bp$ models compared with the PB model, likely associated with band nonparabolicity and warping.\cite{aspnes_band_1974}
A much smaller electroabsorption spectrum is predicted for the DC and optical fields perpendicular compared to fields parallel, as shown in Fig.~\ref{electroabs_even}b for a few polarization combinations.
In contrast with our previous results using 2-band or 8-band models,\cite{wahlstrand_polarization_2011} with the 14-band model we observe a photon-energy-independent positive offset in the perpendicular spectra for $\hbar\omega > E_g/2$.
This offset for perpendicular fields is consistent over all DC field directions simulated and deserves further study.

For the most extreme cases, when all optical fields are polarized along or perpendicular to the DC field, the 14-band calculations generally agree with results using a two-parabolic-bands model.\cite{wahlstrand_polarization_2011}
That is, the change in absorption is maximized when the fields are parallel.
We attribute this polarization dependence to the strong effect of the DC field on the intraband matrix element $\bV_{nn}$, which is proportional to the carrier velocity.
This makes intuitive sense: the velocity is directly modified by a parallel DC field.
This idea was confirmed by performing calculations using an 8-band $\bk \cdot \bp$ model in which the intraband transitions were left out of the sum over intermediate bands; the polarization dependence became much less pronounced.\cite{wahlstrand_polarization_2011}
Since the intraband matrix element is just the velocity of the electron or hole, it is basically the same in the 14 band model near the $\Gamma$ point.

As just discussed, the largest change is predicted for all fields parallel.
Unfortunately, the conventional experimental geometry for electroabsorption or electroreflectance uses a DC field perpendicular to the sample surface and thus perpendicular to the direction of the optical field.\cite{shen_franz--keldysh_1995,weiser_differential_2007}
The only experimental investigation we are aware of on the two-photon FKE uses this geometry and photon energies below the half band gap $E_g/2$ in HgCdTe.\cite{cui_modulation_2008}
Observing the largest effect would require applying a large DC field transverse to the sample surface, which is less common but has been achieved in one-photon FKE experiments.\cite{rehn_transverse_1967,ruff_polarization_1996,wahlstrand_uniform-field_2010}
Another experimental challenge is achieving sufficient spectral resolution to resolve Franz-Keldysh oscillations.
While spectral resolution is trivial in linear spectroscopy, achieving good spectral resolution is not a common concern in the most widely deployed form of two-photon absorption spectroscopy, which is Z-scan.\cite{sheik-bahae_high-sensitivity_1989}
In addition, measurements using Z-scan with a tunable laser probably do not have sufficient signal-to-noise ratio to observe the modification of $\beta$ by a DC field.
Techniques using two pulses have been developed that allow high quality nondegenerate two-photon spectroscopic measurements.\cite{negres_two-photon_2002,negres_experiment_2002}
By properly choosing the bandwidth of the pulses and using a high DC field strength (which increases the period of the Franz-Keldysh oscillations), it should be possible to observe Franz-Keldysh oscillations in the nondegenerate TPA spectrum.
Finally, we can expect that observing many Franz-Keldysh oscillations in two-photon absorption will require an extremely uniform DC field and low temperatures, as it does in one-photon absorption.\cite{weiser_differential_2007}

\subsubsection{Odd FKE}

\begin{figure}
\includegraphics[width=8.5cm]{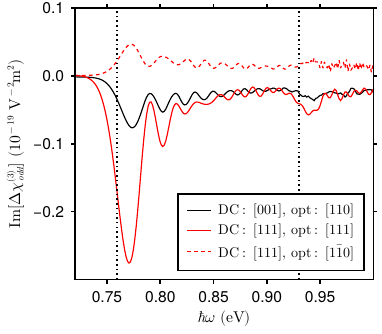}
\caption{Odd two-photon electroabsorption spectra [Eq.~(\ref{oddFKE})] calculated using the 14-band $\bk \cdot \bp$ model in GaAs. All calculations used $\EDC = 44$~kV/cm. Dotted lines indicate $\hbar \omega = E_g/2$ and $\hbar\omega = E_{so}/2$.}
\label{electroabs_odd}
\end{figure}

As discussed earlier, the field-induced change in the absorption coefficient can be odd in the DC field for certain orientations of the optical field.
Unlike the 8-band model, the 14-band model exhibits the lack of inversion symmetry in zincblende crystals, and the odd FKE was calculated for one-photon absorption in our prior paper.\cite{wahlstrand_independent-particle_2010}
To lowest order in the DC field, the odd one-photon effect can be related to $\chi^{(2)}$, though the calculated electroabsorption coefficient at a particular photon energy above the band edge does not simply scale linearly with the field but rather oscillates.
It is important to realize, however, that these oscillations are rather difficult to measure in experiments.
Experiments are typically performed on imperfect crystals at finite temperatures with nonuniform DC fields, which tends to damp the Franz-Keldysh oscillations except near band structure critical points.\cite{aspnes_band_1974}
However the odd FKE differs from the even FKE in an important way: the electroabsorption averages to a non-zero value in the presence of scattering processes and nonuniform fields.
For two-photon electroabsorption, the lowest-order odd FKE is related to $\chi^{(4)}$.
For the zincblende structure, the components $\chi^{(4)}_{xxxyz}$ and permutations are non-zero.
For the two-photon carrier injection tensor, the non-zero elements corresponding to this are $\eta^{xyzz}_2$, $\eta^{xxxy}_2$, and $\eta^{zxyz}_2$, as mentioned earlier and shown in Fig.~\ref{offdiag}b for a DC field pointing along [001].
Electroabsorption spectra, calculated using Eq.~(\ref{oddFKE}) for a few other DC field and optical field polarizations, are shown in Fig.~\ref{electroabs_odd}.

The largest effect is predicted for both fields along [111] (solid red line in Fig.~\ref{electroabs_odd}).
The odd FKE is non-zero for some configurations where the DC and optical fields are perpendicular.
Those might be more favorable for experimental observation since it is straightforward to apply a large, uniform DC field perpendicular to a semiconductor surface by using the surface depletion field in doped semiconductors.
In addition, unlike the even FKE for both fields parallel, as explained above the odd FKE averages to a non-zero value over many oscillation periods, which means it may be detectable with wide bandwidth optical pulses and at room temperature, where decoherence damps Franz-Keldysh oscillations.

\subsection{Bichromatic interference control}
\begin{figure}
\includegraphics[width=8.5cm]{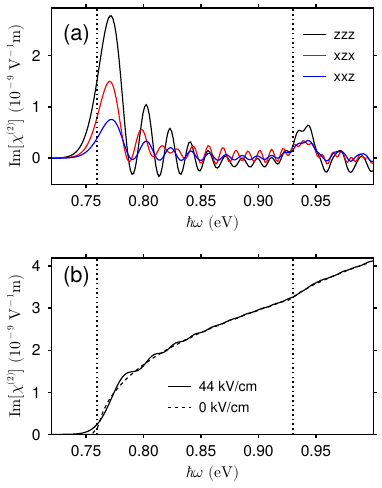}
\caption{Calculation of bichromatic interference control for a DC field along [001]. Dotted lines indicate $\hbar \omega = E_g/2$ and $\hbar\omega = E_{so}/2$. (a) Field-induced elements $zzz$, $xzx$, and $xxz$ for a DC field of 44~kV/cm. (b) Population control element $xyz$ with no DC field (dashed line) and at $\EDC = 44$~kV/cm.}
\label{QUIC}
\end{figure}

There are a number of 1+2-photon (bichromatic) coherent control processes.
As is well known,\cite{atanasov_coherent_1996} a current can be injected by the interference of one-photon absorption at $2\omega$ and two-photon absorption at $\omega$, where $2\hbar\omega$ is above the band gap, with a magnitude dependent on a phase parameter equal to the difference of the phase of the light at $2\omega$ and twice the phase of the light at $\omega$; this is a $\chi^{(3)}$ effect.
In a medium with a $\chi^{(2)}$ and subject to light at $2\omega$ and $\omega$, this interference also leads to a carrier injection that depends on the same phase parameter.\cite{fraser_quantum_1999,fraser_quantum_2003}
This latter process is often called ``population control'' since it allows control of the total injected carrier population via the phase between light field components.
In the absence of a DC field, the only nonzero element of the carrier injection tensor for bichromatic coherent control is $\eta^{xyz}_I$ in a zincblende crystal such as GaAs.
A DC field can also enable population control via bichromatic interference.\cite{wahlstrand_optical_2011,wahlstrand_electric_2011}
The effect depends on the sign of $\EDC$, like the odd FKE, but in this case it is a $\chi^{(3)}$ effect to lowest order, so it does not rely on broken inversion symmetry.
Previous analytical expressions were presented using a two-parabolic-bands model,\cite{wahlstrand_optical_2011} and a few plots were previously shown from a 14-band calculation for a DC field along [001] and [011].\cite{wahlstrand_electric_2011}
Here we present more calculations of field-induced population control using the 14-band model.
When we plot the wavelength dependence of 1+2 population control, we show the imaginary part of the nonlinear susceptibility\cite{rioux_optical_2012,fraser_quantum_1999}
\begin{equation}
  \mathrm{Im}\left[\chi^{(2)}_{ijk}(-2\omega; \omega, \omega) \right] = \frac{\hbar}{3} \eta^{ijk}_I(\omega; \bEDC).
  \end{equation}
For a DC electric field along $\bzhat$, three additional independent elements of the carrier injection tensor become non-zero: $\eta^{xxz}_I$, $\eta^{xzx}_I$, and $\eta^{zzz}_I$, where $x$ and $y$ are interchangeable.
These are shown in Fig.~\ref{QUIC}a for a DC field along [001].
The effect is predicted to be strongest near $E_g/2$.

When the optical fields are aligned along crystal directions such that population control is allowed, the DC field modifies the spectrum of the injection tensor in much the same way that it modifies the one- and two-photon absorption spectra: it enables the process for photon energies below $E_g/2$ and causes oscillations in the spectrum to appear above $E_g/2$, just as in the spectra shown in Figs.~\ref{kp_abs} and \ref{offdiag}a.
The calculated tensor element $\eta^{xyz}_I$ is shown in Fig.~\ref{QUIC}b for a DC field along [100].
The zero field calculation (dashed line) agrees with a previous calculation using the same 14-band model.\cite{bhat_calculations_2006}

\begin{figure}
\includegraphics[width=8.5cm]{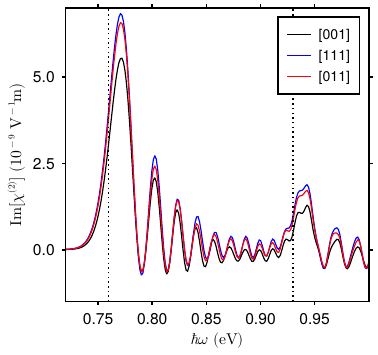}
\caption{Calculation of bichromatic coherent control carrier injection tensor for $\bEDC \parallel \bEopt$ along various crystal directions. All curves used $\EDC = 44$~kV/cm. Dotted lines indicate $\hbar \omega = E_g/2$ and $\hbar\omega = E_{so}/2$.}
\label{EQUIC}
\end{figure}

As seen in Fig.~\ref{QUIC}a and in experiment,\cite{wahlstrand_optical_2011} the largest effect occurs when the DC field and both optical fields (at $\omega$ and $2\omega$) point along the same direction.
Fig.~\ref{EQUIC} shows the injection tensor for parallel fields along various crystal directions.
Minor differences in the Franz-Keldysh oscillation period are due to the slightly different effective mass as a function of direction.\cite{wahlstrand_independent-particle_2010}
Note that in Fig.~\ref{QUIC}a and Fig.~\ref{EQUIC} the carrier injection rate induced by quantum interference becomes negative at certain photon energy values, but the \emph{total} carrier injection rate (including one- and two-photon absorption) remains positive in all cases.
Field-induced bichromatic interference control is predicted to be most effective for the DC field along $\langle 111 \rangle$.

The only experimental results so far on the effect of an electric field on interference control used a fixed excitation wavelength.\cite{wahlstrand_optical_2011,wahlstrand_electric_2011}
Measurements of the wavelength dependence are experimentally challenging because of the need to normalize by the excitation irradiance as the frequencies $\omega$ and $2\omega$ are simultaneously tuned.
Nevertheless it has been done in the case of 1+2-photon population control in the absence of a DC field.\cite{sames_all-optical_2009}

\section{Conclusion and Outlook}
We have presented an independent-particle-approximation theory of the two-photon FKE and demonstrated its use in numerical calculations using a 14-band $\bk \cdot \bp$ model.\cite{pfeffer_five-level_1996}
As in previous calculations using fewer bands,\cite{wahlstrand_polarization_2011} we find a polarization dependence that we attribute to the strong effect of a DC field on intraband dynamics.
We predict changes in the two-photon absorption spectrum that are odd in the DC field that occur because of the lack of inversion symmetry in GaAs.
We also predict changes in the bichromatic population control injection rate in GaAs.

There is still little experimental work in this area, and in the results section we provided brief notes and recommendations for potential experimental observations.
Modern techniques for TPA spectroscopy can achieve sufficient spectral resolution to observe the two-photon FKE with high electric field strength when the two-photon energy is near $E_g$.\cite{negres_two-photon_2002,negres_experiment_2002,fishman_sensitive_2011,cruz_phase-sensitive_2024}
Many considerations for one-photon FKE experiments apply to two-photon experiments.
Measurements at low temperature may be required to observe more than a single Franz-Keldysh oscillation.\cite{kolbe_confinement_1999,weiser_differential_2007}
Achieving a uniform DC field can be especially challenging in the presence of the strong excitation required to observe two-photon absorption because of screening and other effects from photoexcited carriers.
Strong field THz pulses offer another way to achieve sufficient fields but would require adapting the theory to handle time-dependent fields, i.e.~the dynamical FKE.\cite{yacoby_high-frequency_1968,jauho_dynamical_1996,otobe_femtosecond_2016,otobe_time-resolved_2016}

Future extensions to the theory could include adding the Coulomb interaction,\cite{duque-gomez_franzkeldysh_2015} as has been done for previous theories of two-photon absorption,\cite{yang_electric-field_1973-1,kolber_theory_1978,areshev_2-photon_1979} and adding damping/scattering processes, which are expected to reduce the amplitude of the Franz-Keldysh oscillations.\cite{aspnes_band_1974}
The Coulomb interaction particularly affects the electroabsorption lineshape for small fields.\cite{jaeger_excitonic_1998,duque-gomez_franzkeldysh_2015}
Higher-order multi-photon absorption and bichromatic interference\cite{wang_quantum_2019} could be calculated by going to higher order in $\Heff$.
Extensions of the theory could be pursued to calculate the two-photon FKE in semiconductors with an indirect bandgap.\cite{garcia_phonon-assisted_2006}
In the one-photon FKE, the refractive index is modified by a DC field in a similar way as the absorption coefficient.\cite{aspnes_electric_1967}
The analogous effect for the two-photon FKE is field-induced modification of the Kerr coefficient $n_2$.\cite{hutchings_kramers-kronig_1992,hutchings_theory_1995}
This could be calculated from our results using a generalized Kramers-Kronig calculation.\cite{hutchings_kramers-kronig_1992}
Finally, highly nondegenerate multiphoton absorption has been a topic of interest because of its promise for sensitive mid-IR detection.\cite{fishman_sensitive_2011}
That effect is closely related to the multiphoton FKE, and extending the theory to dynamical fields may lead to new insights.


\begin{acknowledgments}
  The authors thank S.~T.~Cundiff for discussions and support.
  J.K.W.~thanks E.~L.~Shirley for helpful suggestions.
  J.E.S.~acknowledges support from the Natural Sciences and Engineering Research Council of Canada, and useful discussions with Jason Kattan.
\end{acknowledgments}


\end{document}


\title{Supplemental document for \\ ``Theory of the two-photon Franz-Keldysh effect \\ and electric field-induced bichromatic coherent control''}
\author{J. K. Wahlstrand}
\affiliation{Nanoscale Device Characterization Division, National Institute of Standards and Technology, Gaithersburg, MD 20899 USA}
\author{J. E. Sipe}
\affiliation{Department of Physics, University of Toronto, Toronto, Ontario, Canada M5S1A7}
\date{\today}

\maketitle

This document includes detailed derivations supporting the theory described in the main text.

\section{$\ket{\Psi^{(2)}}$ in the presence of a DC field \label{AppCalcPsi2}}

Here we derive Eqs.~(7,8) from Eq.~(6).
In Ref.~\onlinecite{wahlstrand_independent-particle_2010}, electron creation and destruction operators $b^\dagger_{v\bk}$ and $b_{v\bk}$ were used.
We begin here by defining hole creation and destruction operators, writing
\begin{math}
p_{v\bk}=b^\dagger_{v\bk}
\end{math}
and
\begin{math}
p^\dagger_{v\bk}=b_{v\bk}.
\end{math}
This leads to
\begin{math}
b^{\dagger}_{v\bk} b_{v'\bk}=p_{v\bk} p^{\dagger}_{v'\bk} = \delta_{vv'}-p^{\dagger}_{v'\bk}p_{v\bk}.
\end{math}
Using this to expand Eq.~(4), we find
\begin{multline}
\tilde{\mathcal{J}} (t) = e\sum_{c,c',\mathbf{k}}b_{c'\mathbf{k}}^{\dagger }b_{c\mathbf{k}}\mathbf{\tilde{V}}_{c'c}(%
\mathbf{k};t) - e\sum_{v,v',\mathbf{k}}p_{v\mathbf{k}}^{\dagger }p_{v'\mathbf{k}}\mathbf{\tilde{V}}_{v'v}(%
\mathbf{k};t) \\
+e\sum_{c,v,\mathbf{k}}b_{c\mathbf{k}}^{\dagger }p^\dagger_{v\mathbf{k}}\mathbf{\tilde{V}}_{cv}(%
\mathbf{k};t) + e\sum_{c,v,\mathbf{k}}p_{v\mathbf{k}} b_{c\mathbf{k}}\mathbf{\tilde{V}}_{vc}(%
\mathbf{k};t).
\label{Jexplicit}
\end{multline}
In Eq.~(6), $H_{\mathrm{eff}}$ is applied to $\left\vert \Psi ^{H}\right\rangle$ twice.
In calculating to first order, only the third term in Eq.~(\ref{Jexplicit}) contributes, because the others yield zero when applied to the vacuum state.
To second order, only the first two terms contribute because they connect conduction bands to conduction bands and valence bands to valence bands.
The fourth term undoes the action of the third term.

With this in mind, the effect of our operator $\mathcal{\tilde{J}}\left( t\right) $
on our Heisenberg ket can be written as 
\begin{equation}
\mathcal{\tilde{J}}\left( t\right) \left\vert \Psi ^{H}\right\rangle =e\sum_{%
\mathbf{k}_{\perp },k\parallel }\sum_{v,c}\mathbf{F}_{cv}(\mathbf{k}_{\perp
};t+\frac{k_{\parallel }}{\varepsilon })\overline{\left\vert cv(\mathbf{k}%
_{\perp }k_{\parallel })\right\rangle },
\label{Jtildeact}
\end{equation}
where $F_{cv}(\bkperp; t)$ is given by Eq.~(10) in the main text.
Using Eq.~(\ref{Jexplicit}) in Eq.~(\ref{Jtildeact}) in Eq.~(2), we have
\begin{equation}
\int_{-\infty}^{t'} \Heff(t^{\prime \prime})\left\vert \Psi ^{H}\right\rangle dt'' = -\frac{e}{c} \sum_{%
\mathbf{k}_{\perp },k_\parallel }\sum_{v,c} \int_{-\infty}^{t'} \mathbf{F}_{cv}(\mathbf{k}_{\perp
};t''+\frac{k_{\parallel }}{\varepsilon }) \cdot \bAopt(t'') \overline{\left\vert cv(\mathbf{k}%
_{\perp }k_{\parallel })\right\rangle } dt''.
\end{equation}
Fourier expanding the optical pulse $\mathbf{E}_{\mathrm{opt}} = -(1/c) \partial\mathbf{A}_\mathrm{opt}(t)/\partial t$ using
\begin{equation}
\bAopt(t)=\int \frac{d\omega }{2\pi }\mathbf{A}(\omega
)e^{-i\omega t},  \label{AFTnew}
\end{equation}
and
\begin{equation}
\mathbf{F}_{cv}(\mathbf{k}_{\perp};t''+\frac{k_{\parallel }}{\varepsilon }) = \int\frac{d\omega }{2\pi }\mathbf{F}_{cv} (\bkperp; -\omega)e^{i\omega t''} e^{i\omega k_\parallel/\varepsilon},
\label{FiFT}
\end{equation}
we find
\begin{eqnarray}
\int_{-\infty}^{t'} \Heff(t^{\prime \prime})\left\vert \Psi ^{H}\right\rangle dt'' &=& 
-\frac{ie}{c} \sum_{\mathbf{k}_{\perp },k_\parallel }\sum_{v,c} \iint \frac{d\omega d\hat{\omega}_1}{4\pi^2} \mathbf{F}_{cv}(\bkperp;-\hat{\omega}_1) \cdot \bAopt(\omega) \frac{e^{-i(\omega-\hat{\omega}_1)t'}}{\omega-\hat{\omega}_1} \nonumber \\ &&\times e^{i\hat{\omega}_1 k_\parallel/\varepsilon}\overline{\left\vert cv(\mathbf{k}_{\perp }k_{\parallel })\right\rangle }.
\label{j99}
\end{eqnarray}

To second order, only the first two terms of Eq.~(\ref{Jexplicit}) contribute.
Expanding using Eq.~(5), we find for these two terms, using the block diagonal approximation to limit the sums over $c,c',c'',c'''$ to the conduction bands and $v,v',v'',v'''$ to the valence bands,
\begin{multline}
\mathcal{\tilde{J}}_2 \left( t\right) = e\sum_{\mathbf{k}_{\perp },k_\parallel} \left( \sum_{c,c',c'',c'''}b_{c'\mathbf{k}_{\perp },k_{\parallel
}}^{\dagger }b_{c\mathbf{k}_{\perp }k_{\parallel }}L_{c''c'}^{\ast }(\mathbf{k}_{\perp },k_{\parallel };t\mathbf{)V}_{c''c'''}(\mathbf{k}_{\perp
},k_{\parallel };t)L_{c'''c}(\mathbf{k}_{\perp },k_{\parallel };t\mathbf{)} \right. \\
\left. -\sum_{v,v',v'',v'''}p_{v\mathbf{k}_{\perp },k_{\parallel
}}^{\dagger }p_{v'\mathbf{k}_{\perp }k_{\parallel }}L_{v''v'}^{\ast }(%
\mathbf{k}_{\perp },k_{\parallel };t\mathbf{)V}_{v''v'''}(\mathbf{k}_{\perp
},k_{\parallel };t)L_{v'''v}(\mathbf{k}_{\perp },k_{\parallel };t\mathbf{)} \right),
\end{multline}
where the subscript 2 signifies that this is the second application of $\mathcal{\tilde{J}}(t)$ to the ket.

One can show that the matrix \textrm{L} can be written as \cite{wahlstrand_independent-particle_2010}
\begin{eqnarray}
L_{pn}(\mathbf{k}_{\perp },k_{\parallel };t)
&=&e^{-i\sigma _{pn}(\mathbf{k}_{\perp },k_{\parallel })}\sum_{q}m_{pq}(%
\mathbf{k}_{\perp };t+\frac{k_{\parallel }}{\varepsilon })\mathcal{B}_{qn}(%
\mathbf{k}_{\perp },k_{\parallel }) \notag  \\
&=&\sum_{q}\bar{m}_{pq}(\mathbf{k}_{\perp };t+\frac{k_{\parallel }}{%
\varepsilon })\mathcal{\bar{B}}_{qn}(\mathbf{k}_{\perp },k_{\parallel }), 
 \label{uuse}
\end{eqnarray}
where
\begin{eqnarray}
\bar{m}_{pq}(\mathbf{k}_{\perp };t+\frac{k_{\parallel }}{\varepsilon })
&=&e^{-i\sigma _{pq}(\mathbf{k}_{\perp },k_{\parallel })}m_{pq}(\mathbf{k}%
_{\perp };t+\frac{k_{\parallel }}{\varepsilon }), \notag \\
\mathcal{\bar{B}}_{qn}(\mathbf{k}_{\perp },k_{\parallel }) &=&e^{-i\sigma
_{qn}(\mathbf{k}_{\perp },k_{\parallel })}\mathcal{B}_{qn}(\mathbf{k}_{\perp
},k_{\parallel }),   \label{barunbar}
\end{eqnarray} and $\bF$ is given by Eq.~(10) in the main text. Using this we have
\begin{multline}
\mathcal{\tilde{J}}_2 \left( t\right)
=e\sum_{\mathbf{k}_{\perp },k\parallel }\sum_{c^{\prime \prime },c'''}\sum_{q,q'}\bar{m}_{c''q'}^{\ast }(\mathbf{k}_{\perp };t+\frac{k_{\parallel }}{\varepsilon })\mathbf{%
V}_{c''c'''}(\mathbf{k}_{\perp },k_{\parallel };t)\bar{m}%
_{c'''q}(\mathbf{k}_{\perp };t+\frac{k_{\parallel }}{%
\varepsilon }) \\
\times \left( \sum_{c'}\mathcal{\bar{B}}_{q'c'}^{\ast }(%
\mathbf{k}_{\perp },k_{\parallel })b_{c'\mathbf{k}_{\perp },k_{\parallel
}}^{\dagger }\right) \left( \sum_{c}\mathcal{\bar{B}}_{qc}(%
\mathbf{k}_{\perp },k_{\parallel })b_{c\mathbf{k}_{\perp }k_{\parallel
}}\right) \\
-e\sum_{\mathbf{k}_{\perp },k\parallel }\sum_{v^{\prime \prime },v'''}\sum_{q,q'}\bar{m}_{v''q'}^{\ast }(\mathbf{k}_{\perp };t+\frac{k_{\parallel }}{\varepsilon })\mathbf{%
V}_{v''v'''}(\mathbf{k}_{\perp },k_{\parallel };t)\bar{m}%
_{v'''q}(\mathbf{k}_{\perp };t+\frac{k_{\parallel }}{%
\varepsilon }) \\
\times \left( \sum_{v'}\mathcal{\bar{P}}_{q'v'}(%
\mathbf{k}_{\perp },k_{\parallel })p_{v'\mathbf{k}_{\perp },k_{\parallel
}}\right) \left( \sum_{v}\mathcal{\bar{P}}^{\ast }_{qv}(%
\mathbf{k}_{\perp },k_{\parallel })p^{\dagger}_{v\mathbf{k}_{\perp }k_{\parallel
}}\right),
\end{multline}
where we have defined
$\mathcal{\bar{P}}_{mn} (\mathbf{k}_{\perp },k_{\parallel }) = \mathcal{\bar{B}}^*_{mn} (\mathbf{k}_{\perp },k_{\parallel })$.

So we have 
\begin{multline}
\mathcal{\tilde{J}}_2\left( t\right)   \label{Jtildeexpand}
=e\sum_{\mathbf{k}_{\perp },k\parallel }\sum_{c'',c'''}\sum_{q,q'}\bar{m}_{c''q'}^{\ast }(\mathbf{k}_{\perp };t+\frac{k_{\parallel }}{\varepsilon })\mathbf{%
V}_{c''c'''}(\mathbf{k}_{\perp },k_{\parallel };t)\bar{m}%
_{c'''q}(\mathbf{k}_{\perp };t+\frac{k_{\parallel }}{%
\varepsilon })B_{q'\mathbf{k}_{\perp },k_{\parallel
}}^{\dagger }B_{q\mathbf{k}_{\perp }k_{\parallel }}
\\
-e\sum_{\mathbf{k}_{\perp },k\parallel }\sum_{v'',v'''}\sum_{q,q'}\bar{m}_{v''q'}^{\ast }(\mathbf{k}_{\perp };t+\frac{k_{\parallel }}{\varepsilon })\mathbf{%
V}_{v''v'''}(\mathbf{k}_{\perp },k_{\parallel };t)\bar{m}%
_{v'''q}(\mathbf{k}_{\perp };t+\frac{k_{\parallel }}{%
\varepsilon })P_{q\mathbf{k}_{\perp },k_{\parallel
}}^{\dagger }P_{q'\mathbf{k}_{\perp }k_{\parallel }}
\\
=e\sum_{\mathbf{k}_{\perp },k\parallel }\sum_{q,q'}e^{-i\sigma _{q'q}(\mathbf{k}%
_{\perp },k_{\parallel })}\mathbf{F}_{q'q}(%
\mathbf{k}_{\perp };t+\frac{k_{\parallel }}{\varepsilon })B_{q'\mathbf{k}_{\perp },k_{\parallel }}^{\dagger }B_{q\mathbf{k}_{\perp }k_{\parallel }} 
\\
-e\sum_{\mathbf{k}_{\perp },k\parallel }\sum_{q,q'}e^{-i\sigma _{q'q}(\mathbf{k}%
_{\perp },k_{\parallel })}\mathbf{F}_{q'q}(%
\mathbf{k}_{\perp };t+\frac{k_{\parallel }}{\varepsilon })P_{q\mathbf{k}_{\perp },k_{\parallel }}^{\dagger }P_{q'\mathbf{k}_{\perp }k_{\parallel }},
\end{multline}
where in the second line we have used Eq.~(\ref{barunbar}).
Using this in the second application of $\Heff(t')$ to Eq.~(\ref{j99}), we find
\begin{multline}
\Heff(t') \int_{-\infty}^{t'} \Heff(t^{\prime \prime})\left\vert \Psi ^{H}\right\rangle dt'' = \\
\frac{ie^2}{c^2} \sum_{\mathbf{k}_{\perp }^\prime,k_\parallel^\prime } \sum_{\mathbf{k}_{\perp },k_\parallel } \sum_{v,c,c',c''} e^{-i\sigma _{c''c'}(\bkperp^\prime,k_{\parallel }^\prime)} \bF_{c''c'}(\bkperp';t'+\frac{k_\parallel^\prime}{\varepsilon}) \cdot \bAopt(t') B_{c''\bkperp',k^\prime_\parallel}^\dagger B_{c'\bkperp',k^\prime_\parallel} \times \\ \iint \frac{d\omega d\hat{\omega}_1}{4\pi^2} \mathbf{F}_{cv}(\bkperp;-\hat{\omega}_1) \cdot \bAopt(\omega) \frac{e^{-i(\omega-\hat{\omega}_1)t'}}{\omega-\hat{\omega}_1} e^{i\hat{\omega}_1 k_\parallel/\varepsilon}\overline{\left\vert cv(\mathbf{k}%
_{\perp }k_{\parallel })\right\rangle } \\
-\frac{ie^2}{c^2} \sum_{\mathbf{k}_{\perp }^\prime,k_\parallel^\prime } \sum_{\mathbf{k}_{\perp },k_\parallel } \sum_{v,v',v'',c} e^{-i\sigma _{v''v'}(\bkperp^\prime,k_{\parallel }^\prime)} \bF_{v''v'}(\bkperp';t'+\frac{k_\parallel^\prime}{\varepsilon}) \cdot \bAopt(t') P_{v'\bkperp',k^\prime_\parallel}^\dagger P_{v''\bkperp',k^\prime_\parallel} \times \\ \iint \frac{d\omega d\hat{\omega}_1}{4\pi^2} \mathbf{F}_{cv}(\bkperp;-\hat{\omega}_1) \cdot \bAopt(\omega) \frac{e^{-i(\omega-\hat{\omega}_1)t'}}{\omega-\hat{\omega}_1} e^{i\hat{\omega}_1 k_\parallel/\varepsilon}\overline{\left\vert cv(\mathbf{k}%
_{\perp }k_{\parallel })\right\rangle }.
\label{j100}
\end{multline}

We next examine the action of the operators.  We find
\begin{multline}
e^{-i\sigma _{q'q}(\bkperp',k_{\parallel }^\prime)} B_{q'\bkperp',k^\prime_\parallel}^\dagger B_{q\bkperp',k^\prime_\parallel} \overline{\left\vert cv(\mathbf{k}_{\perp }k_{\parallel })\right\rangle }
\\
= e^{-i\sigma _{q'q}(\bkperp^\prime,k_{\parallel }^\prime)} e^{-i\sigma _{cv}(\bkperp,k_{\parallel })} B_{q'\bkperp^\prime,k_\parallel^\prime}^\dagger \delta_{qc} \delta_{\bkperp\bkperp'} \delta_{k_\parallel k_\parallel^\prime} P^\dagger_{v\bkperp,k_\parallel} \left \vert \Psi^H \right\rangle,
\label{sigma_c1}
\end{multline}
and
\begin{multline}
e^{-i\sigma _{q'q}(\bkperp',k_{\parallel }^\prime)} P_{q\bkperp',k^\prime_\parallel}^\dagger P_{q'\bkperp',k^\prime_\parallel} \overline{\left\vert cv(\mathbf{k}_{\perp }k_{\parallel })\right\rangle } 
\\
= e^{-i\sigma _{q'q}(\bkperp^\prime,k_{\parallel }^\prime)} e^{-i\sigma _{cv}(\bkperp,k_{\parallel })} B^\dagger_{c\bkperp,k_\parallel} P_{q\bkperp^\prime,k_\parallel^\prime}^\dagger \delta_{q'v} \delta_{\bkperp\bkperp'} \delta_{k_\parallel k_\parallel^\prime}  \left \vert \Psi^H \right\rangle, \\
\label{sigma_v1}
\end{multline}
where we have used the anticommutation relations for $B$ (see Appendix C of Ref.~\onlinecite{wahlstrand_independent-particle_2010}).
Applying Eq.~(\ref{sigma_c1}) and Eq.~(\ref{sigma_v1}) to Eq.~(\ref{j100}) and using $\delta_{\bkperp\bkperp'}$ and $\delta_{k_\parallel k_\parallel'}$ to perform the sums over $\bkperp'$, $k_\parallel'$, $c'$, and $v''$, we find
\begin{multline}
\Heff(t') \int_{-\infty}^{t'} \Heff(t^{\prime \prime})\left\vert \Psi ^{H}\right\rangle dt'' = \\
\frac{ie^2}{c^2}  \sum_{\mathbf{k}_{\perp },k_\parallel } \sum_{v,c,c''}  \bF_{c''c}(\bkperp;t'+\frac{k_\parallel}{\varepsilon}) \cdot \bAopt(t')  \iint \frac{d\omega d\hat{\omega}_1}{4\pi^2} \mathbf{F}_{cv}(\bkperp;-\hat{\omega}_1) \cdot \bAopt(\omega) \\ \times \frac{e^{-i(\omega-\hat{\omega}_1)t'}}{\omega-\hat{\omega}_1} e^{i\hat{\omega}_1 k_\parallel/\varepsilon} \overline{\left\vert c''v(\mathbf{k}_{\perp }k_{\parallel })\right\rangle} \\
-\frac{ie^2}{c^2}  \sum_{\mathbf{k}_{\perp },k_\parallel } \sum_{v,v',c}  \bF_{vv'}(\bkperp;t'+\frac{k_\parallel}{\varepsilon}) \cdot \bAopt(t')  \iint \frac{d\omega d\hat{\omega}_1}{4\pi^2} \mathbf{F}_{cv}(\bkperp;-\hat{\omega}_1) \cdot \bAopt(\omega) \\ \times \frac{e^{-i(\omega-\hat{\omega}_1)t'}}{\omega-\hat{\omega}_1} e^{i\hat{\omega}_1 k_\parallel/\varepsilon} \overline{\left\vert cv'(\mathbf{k}_{\perp }k_{\parallel })\right\rangle},
\end{multline}
where we have used
\begin{math}
\sigma _{pn}(\mathbf{k}_{\perp },k_{\parallel })=\sigma _{pm}(\mathbf{k}%
_{\perp },k_{\parallel })+\sigma _{mn}(\mathbf{k}_{\perp },k_{\parallel })
\end{math}
and Eq.~(C5) of Ref.~\onlinecite{wahlstrand_independent-particle_2010}.

Next we change dummy indices and change $\omega$ to $\omega_1$, yielding
\begin{multline}
\Heff(t') \int_{-\infty}^{t'} \Heff(t^{\prime \prime})\left\vert \Psi ^{H}\right\rangle dt'' = \\
\frac{ie^2}{c^2}  \sum_{\mathbf{k}_{\perp },k_\parallel } \sum_{v,c,c'}  \bF_{cc'}(\bkperp;t'+\frac{k_\parallel}{\varepsilon}) \cdot \bAopt(t') \iint \frac{d\omega_1 d\hat{\omega}_1}{4\pi^2} \mathbf{F}_{c'v}(\bkperp;-\hat{\omega}_1) \cdot \bAopt(\omega_1) \\ \times \frac{e^{-i(\omega_1-\hat{\omega}_1)t'}}{\omega_1-\hat{\omega}_1} e^{i\hat{\omega}_1 k_\parallel/\varepsilon} \overline{\left\vert cv(\mathbf{k}_{\perp }k_{\parallel })\right\rangle} \\
-\frac{ie^2}{c^2}  \sum_{\mathbf{k}_{\perp },k_\parallel } \sum_{v,v',c}  \bF_{v'v}(\bkperp;t'+\frac{k_\parallel}{\varepsilon}) \cdot \bAopt(t') \iint \frac{d\omega_1 d\hat{\omega}_1}{4\pi^2} \mathbf{F}_{cv'}(\bkperp;-\hat{\omega}_1) \cdot \bAopt(\omega_1) \\ \times \frac{e^{-i(\omega_1-\hat{\omega}_1)t'}}{\omega_1-\hat{\omega}_1} e^{i\hat{\omega}_1 k_\parallel/\varepsilon} \overline{\left\vert cv(\mathbf{k}_{\perp }k_{\parallel })\right\rangle}.
\end{multline}
Now we use Eq.~(\ref{AFTnew}) and Eq.~(\ref{FiFT}) to Fourier transform the quantities with $t'$ in the arguments and then integrate over $t'$, so that we have
\begin{multline}
\int_{-\infty}^\infty \Heff(t') \int_{-\infty}^{t'} \Heff(t^{\prime \prime})\left\vert \Psi ^{H}\right\rangle dt'' \\
=\frac{ie^2}{c^2}  \sum_{\mathbf{k}_{\perp },k_\parallel } \sum_{v,c,c'}  \iint \frac{d\omega_1 d\hat{\omega}_1 d\omega_2 d\hat{\omega}_2}{8\pi^3} \bF_{cc'}(\bkperp;-\hat{\omega}_2) \cdot \bAopt(\omega_2) \mathbf{F}_{c'v}(\bkperp;-\hat{\omega}_1) \cdot \bAopt(\omega_1) \\ \times  \frac{\delta(\omega_1+\omega_2-\hat{\omega}_1-\hat{\omega}_2)}{\omega_1-\hat{\omega}_1} e^{i\hat{\omega}_1 k_\parallel/\varepsilon} e^{i\hat{\omega}_2 k_\parallel/\varepsilon} \overline{\left\vert cv(\mathbf{k}_{\perp }k_{\parallel })\right\rangle} \\
-\frac{ie^2}{c^2}  \sum_{\mathbf{k}_{\perp },k_\parallel } \sum_{v,v',c} \iint \frac{d\omega_1 d\hat{\omega}_1d\omega_2 d\hat{\omega}_2}{8\pi^3} \bF_{v'v}(\bkperp;-\hat{\omega}_2) \cdot \bAopt(\omega_2) \mathbf{F}_{cv'}(\bkperp;-\hat{\omega}_1) \cdot \bAopt(\omega_1) \\ \times   \frac{\delta(\omega_1+\omega_2-\hat{\omega}_1-\hat{\omega}_2)}{\omega_1-\hat{\omega}_1} e^{i\hat{\omega}_1 k_\parallel/\varepsilon} e^{i\hat{\omega}_2 k_\parallel/\varepsilon} \overline{\left\vert cv(\mathbf{k}_{\perp }k_{\parallel })\right\rangle}.
\end{multline}

So we have, using $\mathbf{E}(\omega) =i\omega \mathbf{A}(\omega)/c$,
\begin{multline}
\ket{\Psi^{(2)}}
=\frac{ie^2}{\hbar^2}  \sum_{\mathbf{k}_{\perp },k_\parallel } \sum_{v,c,c'}  \iint \frac{d\omega_1 d\hat{\omega}_1 d\omega_2 d\hat{\omega}_2}{8\pi^3 \omega_1 \omega_2} \left[ \bF_{cc'}(\bkperp;-\hat{\omega}_2) \cdot \bEopt(\omega_2) \right] \left[ \mathbf{F}_{c'v}(\bkperp;-\hat{\omega}_1) \cdot \bEopt(\omega_1) \right]  \\  \times \frac{\delta(\omega_1+\omega_2-\hat{\omega}_1-\hat{\omega}_2)}{\omega_1-\hat{\omega}_1} e^{i\hat{\omega}_1 k_\parallel/\varepsilon} e^{i\hat{\omega}_2 k_\parallel/\varepsilon} \overline{\left\vert cv(\mathbf{k}_{\perp }k_{\parallel })\right\rangle} \\
-\frac{ie^2}{\hbar^2}  \sum_{\mathbf{k}_{\perp },k_\parallel } \sum_{v,v',c} \iint \frac{d\omega_1 d\hat{\omega}_1d\omega_2 d\hat{\omega}_2}{8\pi^3\omega_1 \omega_2} \left[ \bF_{v'v}(\bkperp;-\hat{\omega}_2) \cdot \bEopt(\omega_2) \right] \left[ \mathbf{F}_{cv'}(\bkperp;-\hat{\omega}_1) \cdot \bEopt(\omega_1) \right]  \\ \times  \frac{\delta(\omega_1+\omega_2-\hat{\omega}_1-\hat{\omega}_2)}{\omega_1-\hat{\omega}_1} e^{i\hat{\omega}_1 k_\parallel/\varepsilon} e^{i\hat{\omega}_2 k_\parallel/\varepsilon} \overline{\left\vert cv(\mathbf{k}_{\perp }k_{\parallel })\right\rangle}.
\end{multline}
We next exchange the dummy variables $\omega_1$ and $\omega_2$, and the dummy variables $\hat{\omega}_1$ and $\hat{\omega}_2$ in the second term.  This yields
\begin{multline}
\ket{\Psi^{(2)}}
=\frac{ie^2}{\hbar^2}  \sum_{\mathbf{k}_{\perp },k_\parallel } \sum_{v,c,c'}  \iint \frac{d\omega_1 d\hat{\omega}_1 d\omega_2 d\hat{\omega}_2}{8\pi^3 \omega_1 \omega_2} \left[ \bF_{cc'}(\bkperp;-\hat{\omega}_2) \cdot \bEopt(\omega_2) \right] \left[ \mathbf{F}_{c'v}(\bkperp;-\hat{\omega}_1) \cdot \bEopt(\omega_1) \right]  \\  \times \frac{\delta(\omega_1+\omega_2-\hat{\omega}_1-\hat{\omega}_2)}{\omega_1-\hat{\omega}_1} e^{i\hat{\omega}_1 k_\parallel/\varepsilon} e^{i\hat{\omega}_2 k_\parallel/\varepsilon} \overline{\left\vert cv(\mathbf{k}_{\perp }k_{\parallel })\right\rangle} \\
-\frac{ie^2}{\hbar^2}  \sum_{\mathbf{k}_{\perp },k_\parallel } \sum_{v,v',c} \iint \frac{d\omega_1 d\hat{\omega}_1d\omega_2 d\hat{\omega}_2}{8\pi^3\omega_1 \omega_2}  \left[ \mathbf{F}_{cv'}(\bkperp;-\hat{\omega}_2) \cdot \bEopt(\omega_2) \right] \left[ \bF_{v'v}(\bkperp;-\hat{\omega}_1) \cdot \bEopt(\omega_1) \right] \\ \times  \frac{\delta(\omega_1+\omega_2-\hat{\omega}_1-\hat{\omega}_2)}{\omega_2-\hat{\omega}_2} e^{i\hat{\omega}_1 k_\parallel/\varepsilon} e^{i\hat{\omega}_2 k_\parallel/\varepsilon} \overline{\left\vert cv(\mathbf{k}_{\perp }k_{\parallel })\right\rangle}.
\label{j110}
\end{multline}

We next change variables, using
$\omega _{a} =(\omega _{1}+\omega _{2})/2$,
$\omega_d =\omega _{1}-\omega _{2}$,
$\hat{\omega}_{a} =(\hat{\omega}_{1}+\hat{\omega}_{2})/2$,
and $\omega_m =\hat{\omega}_{1}-\hat{\omega}_{2}$.
Now $1/(2\pi \omega _{1}\omega _{2})=2/[\pi (4\omega _{a}^{2}-\omega
_{d}^{2})]$,
while the Dirac delta function becomes $\delta (2\omega _{a}-2\hat{\omega}_{a})$ and so we have $1/(\omega _{1}-\hat{\omega}_{1}) = 2/(\omega _{d}-\omega_m)$, and $1/(\omega _{2}-\hat{\omega}_{2}) =-2/(\omega _{d}-\omega_m)$.
Also note that the absolute value of the determinant of the Jacobian is 1, so there are no additional factors as we change variables.  The sign of the second term changes, so we can combine the two terms and sum over all intermediate states $n$.

We now have
\begin{multline}
\ket{\Psi^{(2)}}
=\frac{ie^2}{\hbar^2}  \sum_{\mathbf{k}_{\perp },k_\parallel } \sum_{v,c,n}  \iint \frac{d\omega_a d\hat{\omega}_a d\omega_d d\omega_m}{\pi^3 (4\omega_a^2-\omega_d^2)} \left[ \bF_{cn}(\bkperp;-\hat{\omega}_a+\frac{1}{2}\omega_m) \cdot \bEopt(\omega_a-\frac{1}{2}\omega_d) \right] \\ \times\left[ \mathbf{F}_{nv}(\bkperp;-\hat{\omega}_a-\frac{1}{2}\omega_m) \cdot \bEopt(\omega_a+\frac{1}{2}\omega_d) \right]\frac{\delta(2\omega_a-2\hat{\omega}_a)}{\omega_d-\omega_m} e^{2i\hat{\omega}_a k_\parallel/\varepsilon} \overline{\left\vert cv(\mathbf{k}_{\perp }k_{\parallel })\right\rangle}.
\label{j111}
\end{multline}
We can now do the integral over $\hat{\omega}_{a}$, recalling that $\delta
\left( 2\omega _{a}-2\hat{\omega}_{a}\right) =\delta \left( \omega _{a}-\hat{%
\omega}_{a}\right) /2$, and we find 
\begin{multline}
\ket{\Psi^{(2)}}
=\frac{ie^2}{2\hbar^2}  \sum_{\mathbf{k}_{\perp },k_\parallel } \sum_{v,c,n}  \iiint \frac{d\omega_a d\omega_d d\omega_m}{\pi^3 (4\omega_a^2-\omega_d^2)(\omega_d-\omega_m)} \left[ \bF_{cn}(\bkperp;-\omega_a+\frac{1}{2}\omega_m) \cdot \bEopt(\omega_a-\frac{1}{2}\omega_d) \right] \\ \times\left[ \mathbf{F}_{nv}(\bkperp;-\omega_a-\frac{1}{2}\omega_m) \cdot \bEopt(\omega_a+\frac{1}{2}\omega_d) \right] e^{2i\omega_a k_\parallel/\varepsilon} \overline{\left\vert cv(\mathbf{k}_{\perp }k_{\parallel })\right\rangle},
\label{j112}
\end{multline}
leading to Eq.~(7) in the main text, with
\begin{equation}
\theta _{cv\mathbf{k}_{\perp }}^{ij}(\omega _{a},\omega _{d}) =\frac{ie^{2}}{2\pi^3 \hbar ^{2}(4\omega _{a}^{2}-\omega _{d}^{2})} 
 \sum_{n}\int d\omega_m\frac{\;F_{cn}^{i}(\mathbf{k}_{\perp };-\omega _{a}+\frac{1}{2}\omega_m)F_{nv}^{j}(%
\mathbf{k}_{\perp };-\omega _{a}-\frac{1}{2}\omega_m)}{\omega_{d}-\omega_m }. \nonumber
\end{equation}
Symmetrizing with respect to interchanging $i$ and $j$ and
simultaneously sending $\omega _{d}$ to $-\omega _{d}$ leads to Eq.~(8) in the main text.

\section{Continuous wave limit}

Equations (12) and (15) in the main text can be applied to arbitrary optical pulses through Fourier decomposition.
To calculate carrier injection rate tensors as a function of photon energy, we consider the case of long pulses with a narrow frequency distribution.
Consider pulses of length $T$ with a fixed carrier frequency $\omega_o$.
Writing $\bE = \bE_o e^{-i\omega_o t} + \bE_o^* e^{i\omega_o t}$ for $-T/2 < t < T/2$ and zero otherwise, we find that $E^{i}(\omega +\omega_{d}/2)E^{j}(\omega -\omega _{d}/2)$ can be approximated for $\omega>0$ by
\begin{equation}
E^{i}(\omega +\frac{1}{2}\omega _{d})E^{j}(\omega -\frac{1}{2}\omega _{d})
\rightarrow  E_{o}^{i}E_{o}^{j}\frac{\sin \frac{1}{2}(\omega +\frac{1}{2}%
\omega _{d}-\omega _{o})T}{\frac{1}{2}(\omega +\frac{1}{2}\omega _{d}-\omega
_{o})}\frac{\sin \frac{1}{2}(\omega -\frac{1}{2}\omega _{d}-\omega _{o})T}{%
\frac{1}{2}(\omega -\frac{1}{2}\omega _{d}-\omega _{o})}.
\end{equation}
Clearly $\omega _{d}$ must be close to zero to get a significant contribution.
This leads to
\begin{equation}
\int d\omega _{d}E^{i}(\omega +\frac{1}{2}\omega _{d})E^{j}(\omega -\frac{1%
}{2}\omega _{d}) \rightarrow  4\pi E_{o}^{i}E_{o}^{j}\frac{\sin (\omega _{o}-\omega )T}{%
(\omega _{o}-\omega )} \mathrm{~for~}\omega >0,
\end{equation}
and thus
\begin{equation}
E^{i}(\omega +\frac{1}{2}\omega _{d})E^{j}(\omega -\frac{1}{2}\omega
_{d})=4\pi E_{o}^{i}E_{o}^{j}\delta (\omega _{d})\frac{\sin (\omega
_{o}-\omega )T}{(\omega _{o}-\omega )},  \label{ddelta}
\end{equation}
in the CW limit.
We similarly find
\begin{multline}
E^{i}(\omega +\frac{\omega_d}{2})E^{j}(\omega -\frac{\omega_d}{2})\left( E^{l}(\omega +\frac{\omega_d^\prime}{2})E^{m}(\omega -\frac{\omega_d^{\prime}}{2})\right) ^{\ast } 
= \\ 16\pi ^{2}E_{o}^{i}E_{o}^{j}\left( E_{o}^{l}E_{o}^{m}\right) ^{\ast
}\delta (\omega _{d})\delta (\omega _{d}^{\prime })\frac{\sin ^{2}(\omega
_{o}-\omega )T}{(\omega _{o}-\omega )^{2}}.
\end{multline}%
The largest strength here comes from $\omega $ close to $\omega _{o}$, and by evaluating the integral over frequency, using
\begin{equation}
\int_{-\infty }^{\infty }\frac{\sin ^{2}(\omega -\omega _{o})T}{(\omega
-\omega _{o})^{2}}d\omega =\frac{1}{T}T^{2}\int \frac{\sin ^{2}x}{x^{2}}%
dx=\pi T,  \label{sincintegral2}
\end{equation}
leading to Eq.~(16) in the main text.

For bichromatic coherent control, we can take, for positive $\omega $, 
\[
E_{S}^{i}(\omega )=E_{S}^{i}\frac{\sin \frac{1}{2}(\omega -2\omega _{o})T}{%
\frac{1}{2}(\omega -2\omega _{o})}, 
\]%
and so combining this with Eq.~(\ref{ddelta}) we have
\begin{equation}
E_{F}^{l}(\omega +\frac{1}{2}\omega _{d})E_{F}^{m}(\omega -\frac{1}{2}
\omega _{d})\left( E_{S}^{i}(2\omega )\right) ^{\ast } \rightarrow 4\pi E_{F}^{l}E_{F}^{m}\left( E_{S}^{i}\right) ^{\ast }\delta
(\omega _{d})\frac{\sin ^{2}(\omega _{o}-\omega )T}{(\omega _{o}-\omega )^{2}
}.
\end{equation}
Using Eq.~(\ref{sincintegral2}) we then find, for $\omega>0$, Eq.~(22) in the main text.

\section{Two-photon absorption with no DC field \label{AppTwoPhotonNoField}}
Here we calculate the two-photon carrier injection rate with no DC field.

\subsection{Calculation of $\ket{\Psi^{(2)}}$}
From Eqs.~(32,33) in Ref.~\onlinecite{wahlstrand_independent-particle_2010}, we have 
\begin{equation*}
\Heff(t)=-\frac{e}{c}\sum_{n,q,\mathbf{k}}\mathbf{v}_{nq}(\mathbf{k})\cdot \bAopt(t)e^{i\omega_{nq}(\mathbf{k})t}b_{n\mathbf{k}}^{\dagger }b_{q\mathbf{k}}.
\end{equation*}
Note that as with the DC field case, the $\delta _{vv^{\prime }}$ term will make no contribution in $\Heff(t)$ because the sum of $\mathbf{v}_{nn}(\mathbf{k})$ over an
entire band vanishes.
Further
$b_{c\mathbf{k}}^{\dagger }b_{v\mathbf{k}} = b_{c\mathbf{k}}^{\dagger }p_{v\mathbf{k}}^{\dagger }$, and $b_{v\mathbf{k}}^{\dagger }b_{c\mathbf{k}} =p_{v\mathbf{k}}b_{c\mathbf{k}}$,
so
\begin{multline}
\Heff(t) =-\frac{e}{c} \bAopt(t) \cdot \left( \sum_{c,c^{\prime },\mathbf{k}}\mathbf{v}_{cc^{\prime }}(\mathbf{k}) e^{i\omega_{cc^{\prime}}(\mathbf{k})t}b_{c\mathbf{k}}^{\dagger }b_{c^{\prime }\mathbf{k}}
\label{H1use} \right. \\
-\sum_{v,v^{\prime },\mathbf{k}}\mathbf{v}_{vv^{\prime }}(\mathbf{k}) e^{i\omega_{vv^{\prime }}(\mathbf{k})t}p_{v^{\prime }\mathbf{k}}^{\dagger }p_{v\mathbf{k}} 
+\sum_{c,v,\mathbf{k}}\mathbf{v}_{cv}(\mathbf{k})e^{i\omega_{cv}(\mathbf{k})t}b_{c\mathbf{k}}^{\dagger }p_{v\mathbf{k}}^{\dagger }  \\
\left. +\sum_{c,v,\mathbf{k}}\mathbf{v}_{vc}(\mathbf{k})e^{i\omega_{vc}(\mathbf{k})t}p_{v\mathbf{k}}b_{c\mathbf{k}} \right). 
\end{multline}
Only the third term in Eq.~(\ref{H1use}) contributes to first order; we have
\[
\Heff(t^{\prime \prime })\left\vert \Psi^H \right\rangle =-\frac{e}{c} \sum_{c,v,\mathbf{k}}\mathbf{v}_{cv}(\mathbf{k})\cdot \bAopt(t^{\prime \prime })e^{i\omega_{cv}(\mathbf{k})t^{\prime \prime}}\left\vert cv(\mathbf{k})\right\rangle.
\]

Fourier expanding $\mathbf{A}_{\mathrm{opt}}(t)$ and introducing a factor $e^{\eta t''} = e^{-i(+i\eta t'')}$ to ensure a cut off at $-\infty$, we have\cite{wahlstrand_independent-particle_2010}
\begin{eqnarray}
\int_{-\infty }^{t^{\prime }}\Heff(t^{\prime \prime })\left\vert
\Psi^H\right\rangle dt^{\prime \prime }
&=&-\frac{ie}{c}\sum_{c,v,\mathbf{k}}\int \frac{d\omega }{2\pi }\mathbf{v}%
_{cv}(\mathbf{k})\cdot \mathbf{A}(\omega )\frac{e^{-i(\omega -\omega _{cv}(%
\mathbf{k})+i\eta )t^{\prime }}}{\omega -\omega _{cv}(\mathbf{k})+i\eta }%
\left\vert cv(\mathbf{k})\right\rangle . \label{H1first} 
\end{eqnarray}%
In letting $\Heff(t^{\prime })$ act on this we want to keep the
first two terms in Eq.~(\ref{H1use}). \ The third will lead to a second
electron-hole pair, and the fourth will just take us back to the ground state
to keep normalization in order. \ So for acting on the expression Eq.~(\ref%
{H1first}) we can use 
\begin{eqnarray*}
\Heff(t^{\prime }) &\rightarrow &-\frac{e}{c}\sum_{c_{1},c_{2},\mathbf{%
k}^{\prime }}\mathbf{v}_{c_{1}c_{2}}(\mathbf{k}^{\prime })\cdot \bAopt(t')e^{i\omega _{c_{1}c_{2}}(\mathbf{k}^{\prime })t'}b_{c_{1}\mathbf{k}%
^{\prime }}^{\dagger }b_{c_{2}\mathbf{k}^{\prime }} \\
&&+\frac{e}{c}\sum_{v_{1},v_{2},\mathbf{k}^{\prime }}\mathbf{v}_{v_{1}v_{2}}(%
\mathbf{k}^{\prime })\cdot \bAopt(t')e^{i\omega _{v_{1}v_{2}}(%
\mathbf{k}^{\prime })t'}p_{v_{2}\mathbf{k}^{\prime }}^{\dagger }p_{v_{1}%
\mathbf{k}^{\prime }}.
\end{eqnarray*}%
Now we have 
\begin{eqnarray}
b_{c_{1}\mathbf{k}^{\prime }}^{\dagger }b_{c_{2}\mathbf{k}^{\prime
}}\left\vert cv(\mathbf{k})\right\rangle &=&b_{c_{1}\mathbf{k}^{\prime
}}^{\dagger }b_{c_{2}\mathbf{k}^{\prime }}b_{c\mathbf{k}}^{\dagger }p_{v%
\mathbf{k}}^{\dagger }\left\vert \Psi^H \right\rangle  \label{iden1} \\
&=&\delta _{c_{2}c}\delta _{\mathbf{k}^{\prime }\mathbf{k}}b_{c_{1}\mathbf{k}%
^{\prime }}^{\dagger }p_{v\mathbf{k}}^{\dagger }\left\vert \Psi^H\right\rangle
-b_{c_{1}\mathbf{k}^{\prime }}^{\dagger }b_{c\mathbf{k}}^{\dagger }b_{c_{2}%
\mathbf{k}^{\prime }}p_{v\mathbf{k}}^{\dagger }\left\vert \Psi^H\right\rangle 
\nonumber \\
&=&\delta _{c_{2}c}\delta _{\mathbf{k}^{\prime }\mathbf{k}}\left\vert c_{1}v(%
\mathbf{k})\right\rangle,  \nonumber
\end{eqnarray}%
while 
\begin{eqnarray}
p_{v_{2}\mathbf{k}^{\prime }}^{\dagger }p_{v_{1}\mathbf{k}^{\prime
}}\left\vert cv(\mathbf{k})\right\rangle &=&p_{v_{2}\mathbf{k}^{\prime
}}^{\dagger }p_{v_{1}\mathbf{k}^{\prime }}b_{c\mathbf{k}}^{\dagger }p_{v%
\mathbf{k}}^{\dagger }\left\vert \Psi^H\right\rangle  \label{iden2} \\
&=&-p_{v_{2}\mathbf{k}^{\prime }}^{\dagger }b_{c\mathbf{k}}^{\dagger
}p_{v_{1}\mathbf{k}^{\prime }}p_{v\mathbf{k}}^{\dagger }\left\vert
\Psi^H\right\rangle  \nonumber \\
&=&b_{c\mathbf{k}}^{\dagger }p_{v_{2}\mathbf{k}^{\prime }}^{\dagger }p_{v_{1}%
\mathbf{k}^{\prime }}p_{v\mathbf{k}}^{\dagger }\left\vert \Psi^H\right\rangle 
\nonumber \\
&=&\delta _{v_{1}v}\delta _{\mathbf{k}^{\prime }\mathbf{k}}b_{c\mathbf{k}%
}^{\dagger }p_{v_{2}\mathbf{k}^{\prime }}^{\dagger }\left\vert
\Psi^H\right\rangle -b_{c\mathbf{k}}^{\dagger }p_{v_{2}\mathbf{k}^{\prime
}}^{\dagger }p_{v\mathbf{k}}^{\dagger }p_{v_{1}\mathbf{k}^{\prime
}}\left\vert \Psi^H\right\rangle  \nonumber \\
&=&\delta _{v_{1}v}\delta _{\mathbf{k}^{\prime }\mathbf{k}}\left\vert cv_{2}(%
\mathbf{k})\right\rangle .  \nonumber
\end{eqnarray}
So we have 
\begin{eqnarray*}
&&\Heff(t^{\prime })\int_{-\infty }^{t^{\prime }}\Heff(t^{\prime \prime })\left\vert \Psi^H \right\rangle dt^{\prime \prime } \\
&=&\frac{ie^{2}}{c^{2}}\sum_{c,c_{1},v,\mathbf{k}}\mathbf{v}_{c_{1}c}(\mathbf{k}%
)\cdot \bAopt(t^{\prime })e^{i\omega _{c_{1}c}(\mathbf{k}%
)t^{\prime }}\int \frac{d\omega }{2\pi }\mathbf{v}_{cv}(\mathbf{k})\cdot 
\mathbf{A}(\omega )\frac{e^{-i(\omega -\omega _{cv}(\mathbf{k}))t^{\prime }}%
}{\omega -\omega _{cv}(\mathbf{k})+i\eta}\left\vert c_{1}v(\mathbf{k}%
)\right\rangle \\
&&-\frac{ie^{2}}{c^{2}}\sum_{c,v,v_{2},\mathbf{k}}\mathbf{v}_{vv_{2}}(\mathbf{k}%
)\cdot \bAopt(t^{\prime })e^{i\omega _{vv_{2}}(\mathbf{k}%
)t^{\prime }}\int \frac{d\omega }{2\pi }\mathbf{v}_{cv}(\mathbf{k})\cdot 
\mathbf{A}(\omega )\frac{e^{-i(\omega -\omega _{cv}(\mathbf{k}))t^{\prime }}%
}{\omega -\omega _{cv}(\mathbf{k})+i\eta}\left\vert cv_{2}(\mathbf{k}%
)\right\rangle .
\end{eqnarray*}

Changing dummy indices in both expressions yields
\begin{multline}
\Heff(t^{\prime })\int_{-\infty }^{t^{\prime }}\Heff(t^{\prime \prime })\left\vert \Psi^H \right\rangle dt^{\prime \prime } 
= \\ \frac{ie^{2}}{c^{2}}\sum_{c,c^{\prime },v,\mathbf{k}}\mathbf{v}_{cc^{\prime
}}(\mathbf{k})\cdot \bAopt(t^{\prime })e^{i\omega _{cc^{\prime }}(%
\mathbf{k})t^{\prime }}\int \frac{d\omega }{2\pi }\mathbf{v}_{c^{\prime }v}(%
\mathbf{k})\cdot \mathbf{A}(\omega )\frac{e^{-i(\omega -\omega _{c^{\prime
}v}(\mathbf{k}))t^{\prime }}}{\omega -\omega _{c^{\prime }v}(\mathbf{k}%
)+i\eta }\left\vert cv(\mathbf{k})\right\rangle \\
-\frac{ie^{2}}{c^{2}}\sum_{c,v,v^{\prime },\mathbf{k}}\mathbf{v}_{v^{\prime
}v}(\mathbf{k})\cdot \bAopt(t^{\prime })e^{i\omega _{v^{\prime }v}(%
\mathbf{k})t^{\prime }}\int \frac{d\omega }{2\pi }\mathbf{v}_{cv^{\prime }}(%
\mathbf{k})\cdot \mathbf{A}(\omega )\frac{e^{-i(\omega -\omega _{cv^{\prime
}}(\mathbf{k}))t^{\prime }}}{\omega -\omega _{cv^{\prime }}(\mathbf{k}%
)+i\eta }\left\vert cv(\mathbf{k})\right\rangle .
\end{multline}
Since 
\begin{math}
\omega _{cc^{\prime }}(\mathbf{k})+\omega _{c^{\prime }v}(\mathbf{k})=\omega
_{v^{\prime }v}(\mathbf{k})+\omega _{cv^{\prime }}(\mathbf{k})=\omega _{cv}(%
\mathbf{k}), 
\end{math}
we can write this, Fourier expanding $\mathbf{A}_{\mathrm{opt}} (t')$ in terms of $\omega_2$, as
\begin{eqnarray*}
&&\Heff(t^{\prime })\int_{-\infty }^{t^{\prime }}\Heff(t^{\prime \prime })\left\vert \Psi^H \right\rangle dt^{\prime \prime } \\
&=&\frac{ie^{2}}{c^{2}}\sum_{c,c^{\prime },v,\mathbf{k}}\int \frac{d\omega
_{2}d\omega _{1}}{(2\pi )^{2}}\left( \mathbf{v}_{cc^{\prime }}(\mathbf{k}%
)\cdot \mathbf{A}(\omega _{2})\right) \left( \mathbf{v}_{c^{\prime }v}(%
\mathbf{k})\cdot \mathbf{A}(\omega _{1})\right) \frac{e^{-i(\omega
_{2}+\omega _{1}-\omega _{cv}(\mathbf{k}))t^{\prime }}}{\omega _{1}-\omega
_{c^{\prime }v}(\mathbf{k})+i\eta }\left\vert cv(\mathbf{k})\right\rangle \\
&&-\frac{ie^{2}}{c^{2}}\sum_{c,v,v^{\prime },\mathbf{k}}\int \frac{d\omega
_{2}d\omega _{1}}{(2\pi )^{2}}\left( \mathbf{v}_{v^{\prime }v}(\mathbf{k}%
)\cdot \mathbf{A}(\omega _{2})\right) \left( \mathbf{v}_{cv^{\prime }}(%
\mathbf{k})\cdot \mathbf{A}(\omega _{1})\right) \frac{e^{-i(\omega
_{2}+\omega _{1}-\omega _{cv}(\mathbf{k}))t^{\prime }}}{\omega _{1}-\omega
_{cv^{\prime }}(\mathbf{k})+i\eta }\left\vert cv(\mathbf{k})\right\rangle .
\end{eqnarray*}
The time integral then gives
\begin{multline}
\int_{-\infty }^{\infty }\Heff(t^{\prime })\int_{-\infty}^{t^{\prime }}\Heff(t^{\prime \prime })\left\vert \Psi^H \right\rangle
dt^{\prime \prime } \\
=\frac{ie^{2}}{c^{2}}\sum_{c,c^{\prime },v,\mathbf{k}}\int \frac{d\omega
_{2}d\omega _{1}}{2\pi }\left( \mathbf{v}_{cc^{\prime }}(\mathbf{k})\cdot 
\mathbf{A}(\omega _{2})\right) \left( \mathbf{v}_{c^{\prime }v}(\mathbf{k}%
)\cdot \mathbf{A}(\omega _{1})\right)  \frac{\delta \left( \omega _{2}+\omega _{1}-\omega _{cv}(\mathbf{k}%
)\right) }{\omega _{1}-\omega _{c^{\prime }v}(\mathbf{k})+i\eta }%
\left\vert cv(\mathbf{k})\right\rangle \\
-\frac{ie^{2}}{c^{2}}\sum_{c,v,v^{\prime },\mathbf{k}}\int \frac{d\omega
_{2}d\omega _{1}}{2\pi }\left( \mathbf{v}_{v^{\prime }v}(\mathbf{k})\cdot 
\mathbf{A}(\omega _{2})\right) \left( \mathbf{v}_{cv^{\prime }}(\mathbf{k}%
)\cdot \mathbf{A}(\omega _{1})\right) \frac{\delta \left( \omega _{2}+\omega _{1}-\omega _{cv}(\mathbf{k}%
)\right) }{\omega _{1}-\omega _{cv^{\prime }}(\mathbf{k})+i\eta}%
\left\vert cv(\mathbf{k})\right\rangle .
\end{multline}
Then we can write (in terms of the optical field)
\begin{eqnarray*}
\left\vert \Psi^{(2)}\right\rangle 
&=&\frac{ie^{2}}{\hbar ^{2}}\sum_{c,c^{\prime },v,\mathbf{k}}\int \frac{d\omega
_{2}d\omega _{1}}{2\pi \omega _{1}\omega _{2}}\left( \mathbf{v}_{cc^{\prime
}}(\mathbf{k})\cdot \mathbf{E}(\omega _{2})\right) \left( \mathbf{v}%
_{c^{\prime }v}(\mathbf{k})\cdot \mathbf{E}(\omega _{1})\right) \frac{\delta \left( \omega _{2}+\omega _{1}-\omega _{cv}(\mathbf{k}%
)\right) }{\omega _{1}-\omega _{c^{\prime }v}(\mathbf{k})+i\eta }%
\left\vert cv(\mathbf{k})\right\rangle \\
&&-\frac{ie^{2}}{\hbar ^{2}}\sum_{c,v,v^{\prime },\mathbf{k}}\int \frac{d\omega
_{2}d\omega _{1}}{2\pi \omega _{1}\omega _{2}}\left( \mathbf{v}_{v^{\prime
}v}(\mathbf{k})\cdot \mathbf{E}(\omega _{2})\right) \left( \mathbf{v}%
_{cv^{\prime }}(\mathbf{k})\cdot \mathbf{E}(\omega _{1})\right) \frac{\delta \left( \omega _{2}+\omega _{1}-\omega _{cv}(\mathbf{k}%
)\right) }{\omega _{1}-\omega _{cv^{\prime }}(\mathbf{k})+i\eta }%
\left\vert cv(\mathbf{k})\right\rangle .
\end{eqnarray*}%
The symmetry of the final result will be clearer if we swap the dummy
variables $\omega _{1}$ and $\omega _{2}$ in the second expression, giving 
\begin{multline}
\left\vert \Psi^{(2)}\right\rangle  \label{psiout2use} 
= \\ \frac{ie^{2}}{\hbar ^{2}}\sum_{c,c^{\prime },v,\mathbf{k}}\int \frac{d\omega
_{2}d\omega _{1}}{2\pi \omega _{1}\omega _{2}}\left( \mathbf{v}_{cc^{\prime
}}(\mathbf{k})\cdot \mathbf{E}(\omega _{2})\right) \left( \mathbf{v}%
_{c^{\prime }v}(\mathbf{k})\cdot \mathbf{E}(\omega _{1})\right)  \frac{\delta \left( \omega _{2}+\omega _{1}-\omega _{cv}(\mathbf{k}%
)\right) }{\omega _{1}-\omega _{c^{\prime }v}(\mathbf{k})+i\eta }%
\left\vert cv(\mathbf{k})\right\rangle   \\
-\frac{ie^{2}}{\hbar ^{2}}\sum_{c,v,v^{\prime },\mathbf{k}}\int \frac{d\omega
_{2}d\omega _{1}}{2\pi \omega _{1}\omega _{2}}\left( \mathbf{v}_{cv^{\prime
}}(\mathbf{k})\cdot \mathbf{E}(\omega _{2})\right) \left( \mathbf{v}%
_{v^{\prime }v}(\mathbf{k})\cdot \mathbf{E}(\omega _{1})\right)  \frac{\delta \left( \omega _{2}+\omega _{1}-\omega _{cv}(\mathbf{k}%
)\right) }{\omega _{2}-\omega _{cv^{\prime }}(\mathbf{k})+i\eta }%
\left\vert cv(\mathbf{k})\right\rangle .
\end{multline}

Before proceeding, we examine how this expression behaves as the frequencies 
$\omega _{1}$ and $\omega _{2}$ approach zero. \ To do this, we use the
Dirac delta function to write 
$\omega _{1}-\omega _{c^{\prime }v}(\mathbf{k}) =-\omega _{2}+\omega_{cc^{\prime }}(\mathbf{k})$, and $\omega _{2}-\omega _{cv^{\prime }}(\mathbf{k}) =-\omega _{1}+\omega_{v^{\prime }v}(\mathbf{k})$,
and so we can write Eq.~(\ref{psiout2use}) as%
\begin{eqnarray*}
\left\vert \Psi^{(2)}\right\rangle 
&=&\frac{ie^{2}}{\hbar ^{2}}\sum_{c,c^{\prime },v,\mathbf{k}}\int \frac{d\omega
_{2}d\omega _{1}}{2\pi \omega _{1}\omega _{2}}\left( \mathbf{v}_{cc^{\prime
}}(\mathbf{k})\cdot \mathbf{E}(\omega _{2})\right) \left( \mathbf{v}%
_{c^{\prime }v}(\mathbf{k})\cdot \mathbf{E}(\omega _{1})\right) \\
&&\times \frac{\delta \left( \omega _{2}+\omega _{1}-\omega _{cv}(\mathbf{k}%
)\right) }{\omega _{cc^{\prime }}(\mathbf{k})-\omega _{2}+i\eta }%
\left\vert cv(\mathbf{k})\right\rangle \\
&&-\frac{ie^{2}}{\hbar ^{2}}\sum_{c,v,v^{\prime },\mathbf{k}}\int \frac{d\omega
_{2}d\omega _{1}}{2\pi \omega _{1}\omega _{2}}\left( \mathbf{v}_{cv^{\prime
}}(\mathbf{k})\cdot \mathbf{E}(\omega _{2})\right) \left( \mathbf{v}%
_{v^{\prime }v}(\mathbf{k})\cdot \mathbf{E}(\omega _{1})\right) \\
&&\times \frac{\delta \left( \omega _{2}+\omega _{1}-\omega _{cv}(\mathbf{k}%
)\right) }{\omega _{v^{\prime }v}(\mathbf{k})-\omega _{1}+i\eta }%
\left\vert cv(\mathbf{k})\right\rangle .
\end{eqnarray*}
Now certainly there will be terms for which $\omega _{cc^{\prime }}(\mathbf{k%
})=0$ and for which $\omega _{v^{\prime }v}(\mathbf{k})=0$ (i.e., where $%
c=c^{\prime }$ and $v=v^{\prime }$). \ But we will assume that $\mathbf{E(}%
\omega )$ vanishes as $\omega ^{2}$ as $\omega \rightarrow 0$, so there will
be no divergence because of them. \ Further, we assume that the amplitudes $%
\mathbf{E}\left( \omega \right) $ are such that we do not have significant $%
\mathbf{E}(\omega _{1})$ and $\mathbf{E}(\omega _{2})$ with both $\omega
_{2}+\omega _{1}=\omega _{cv}(\mathbf{k})$ and $\omega _{2}=\omega
_{cc^{\prime }}(\mathbf{k})$ for $c\neq c^{\prime }$, and we do not have
significant $\mathbf{E}(\omega _{1})$ and $\mathbf{E}(\omega _{2})$ with
both $\omega _{2}+\omega _{1}=\omega _{cv}(\mathbf{k})$ and $\omega
_{2}=\omega _{vv^{\prime }}(\mathbf{k})$ for $v\neq v^{\prime }$. \ These
conditions would correspond to true ``double transitions.'' \ In their
absence, the expression for $\left\vert \Psi^{(2)}\right\rangle $ is
well-behaved. Thus we can drop the $i\eta $ factors.

We return to Eq.~(\ref{psiout2use}) and change frequency arguments, putting
\begin{eqnarray*}
\omega _{a} &=&\frac{\omega _{1}+\omega _{2}}{2}, \\
\omega _{d} &=&\omega _{1}-\omega _{2}.
\end{eqnarray*}%
Then $d\omega _{a}d\omega _{d}=d\omega _{1}d\omega _{2}$, and 
\begin{eqnarray*}
\omega _{1} &=&\omega _{a}+\frac{1}{2}\omega _{d}, \\
\omega _{2} &=&\omega _{a}-\frac{1}{2}\omega _{d}.
\end{eqnarray*}%
Then, since the delta function guarantees that $2\omega _{a}=\omega _{cv}(%
\mathbf{k})$, we have
$\omega _{1}\omega _{2} =\omega _{a}^{2}-\omega _{d}^{2}/4 =(\omega _{cv}^{2}(\mathbf{k})-\omega _{d}^{2})/4$,
so 
$1/(2\pi \omega _{1}\omega _{2})=2/[\pi \left( \omega _{cv}^{2}(\mathbf{k})-\omega _{d}^{2}\right)]$.
Further 
\begin{eqnarray*}
\omega _{1}-\omega _{c^{\prime }v}(\mathbf{k}) &=&
\omega _{a}+\frac{1}{2}\omega _{d}-\omega _{c^{\prime }v}(\mathbf{k}) \\
&=&\frac{1}{2}\left( \omega _{cc^{\prime }}(\mathbf{k})+\omega _{vc^{\prime
}}(\mathbf{k})+\omega _{d}\right),
\end{eqnarray*}%
while%
\begin{eqnarray*}
\omega _{2}-\omega _{cv^{\prime }}(\mathbf{k}) &=&\omega _{a}-\frac{1}{2}\omega _{d}-\omega _{cv^{\prime }}(\mathbf{k}) \\
&=&-\frac{1}{2}\left( \omega _{cv^{\prime }}(\mathbf{k})+\omega _{vv^{\prime
}}(\mathbf{k})+\omega _{d}\right).
\end{eqnarray*}
With this we can rewrite our expression for $\left\vert \Psi^{(2)}\right\rangle$ as an unrestricted sum over conduction and valence bands $n$,
\begin{eqnarray*}
\left\vert \Psi^{(2)}\right\rangle 
&=&\frac{4ie^{2}}{\pi \hbar ^{2}}\sum_{c,n,v,\mathbf{k}}\int \frac{d\omega
_{a}d\omega _{d}}{\omega _{cv}^{2}(\mathbf{k})-\omega _{d}^{2}}\left( 
\mathbf{v}_{cn}(\mathbf{k})\cdot \mathbf{E}(\omega _{a}-\frac{1}{2}\omega
_{d})\right) \left( \mathbf{v}_{nv}(\mathbf{k})\cdot \mathbf{E}(\omega _{a}+%
\frac{1}{2}\omega _{d})\right) \\
&&\times \frac{\delta \left( 2\omega _{a}-\omega _{cv}(\mathbf{k})\right) }{%
 \omega _{cn}(\mathbf{k})+\omega _{vn}(\mathbf{k})+\omega _{d} }%
\left\vert cv(\mathbf{k})\right\rangle.
\end{eqnarray*}%
We can then write 
\begin{multline}
\left\vert \Psi^{(2)}\right\rangle  \label{psi2noDC} 
=\sum_{c,v,\mathbf{k}}\int d\omega _{a}d\omega _{d}\;\gamma _{cv\mathbf{k}%
}^{ij}(\omega _{d})\;E^{i}(\omega _{a}-\frac{1}{2}\omega _{d})E^{j}(\omega
_{a}+\frac{1}{2}\omega _{d})\;\delta \left( 2\omega _{a}-\omega _{cv}(%
\mathbf{k})\right) \;\left\vert cv(\mathbf{k})\right\rangle ,
\end{multline}
where we provisionally put%
\[
\gamma _{cv\mathbf{k}}^{ij}(\omega _{d})=\frac{4ie^{2}}{\pi \hbar
^{2}(\omega _{cv}^{2}(\mathbf{k})-\omega _{d}^{2})}\sum_{n}\frac{v_{cn}^{i}(%
\mathbf{k})v_{nv}^{j}(\mathbf{k})}{\omega _{cn}(\mathbf{k})+\omega
_{vn}(\mathbf{k})+\omega _{d} }. 
\]%
Now since changing $i\rightarrow j$ and $\omega _{d}\rightarrow -\omega _{d}$
in our expression will not change any of the final results, it is convenient
to define $\gamma _{cv\mathbf{k}}^{ij}(\omega _{d})$ to satisfy the property
that it is unchanged upon changing $i\rightarrow j$ and $\omega
_{d}\rightarrow -\omega _{d}$. Symmetrizing it this way, we find Eq.~(20) in the main text.

\subsection{Injected carriers}

We now calculate the number of carriers injected:%
\begin{multline}
\Delta N =\left\langle \Psi^{(2)}|\Psi^{(2)}\right\rangle
=\sum_{c,v,\mathbf{k}}\int d\omega _{a}d\omega _{d}d\omega _{a}^{\prime
}d\omega _{d}^{\prime }\;\left( \gamma _{cv\mathbf{k}}^{ij}(\omega
_{d})\right) \left( \gamma _{cv\mathbf{k}}^{km}(\omega _{d}^{\prime
})\right) ^{\ast }\; \\
\times E^{i}(\omega _{a}-\frac{1}{2}\omega _{d})E^{j}(\omega _{a}+\frac{1}{%
2}\omega _{d})\;\left( E^{l}(\omega _{a}^{\prime }-\frac{1}{2}\omega
_{d}^{\prime })E^{m}(\omega _{a}^{\prime }+\frac{1}{2}\omega _{d}^{\prime
})\right) ^{\ast } \\ \times \delta \left( 2\omega _{a}-\omega _{cv}(\mathbf{k})\right) \delta
\left( 2\omega _{a}^{\prime }-\omega _{cv}(\mathbf{k})\right) , \notag
\end{multline}%
where we have used the fact that the states $\left\vert cv(\mathbf{k}%
)\right\rangle $ are orthonormal. Now integrating over $\omega _{a}^{\prime
} $ and using $\delta (2\omega _{a}^{\prime }-\omega _{cv}(\mathbf{k}%
))=\delta (\omega _{a}^{\prime }-\omega _{cv}(\mathbf{k})/2)/2$, we replace $%
\omega _{a}^{\prime }$ by $\omega _{cv}(\mathbf{k})/2$ which then, by the
other Dirac delta function, can be replaced by $\omega _{a}$. So we have 
\begin{eqnarray*}
\Delta N 
&=&\sum_{c,v,\mathbf{k}}\int \frac{d\omega _{a}d\omega _{d}d\omega
_{d}^{\prime }}{2}\;\left( \gamma _{cv\mathbf{k}}^{ij}(\omega _{d})\right)
\left( \gamma _{cv\mathbf{k}}^{km}(\omega _{d}^{\prime })\right) ^{\ast } \\
&&\times E^{i}(\omega _{a}-\frac{1}{2}\omega _{d})E^{j}(\omega _{a}+\frac{1}{%
2}\omega _{d})\;\left( E^{l}(\omega _{a}-\frac{1}{2}\omega _{d}^{\prime
})E^{m}(\omega _{a}+\frac{1}{2}\omega _{d}^{\prime })\right) ^{\ast }\delta
\left( 2\omega _{a}-\omega _{cv}(\mathbf{k})\right) .
\end{eqnarray*}%
Converting the sums over $\mathbf{k}$ to volume integrals, we have 
\begin{eqnarray*}
\Delta n 
&=&\int d\omega _{a}d\omega _{d}d\omega _{d}^{\prime }\left( \sum_{c,v}\int 
\frac{d\mathbf{k}}{8\pi ^{3}}\frac{1}{2}\left( \gamma _{cv\mathbf{k}%
}^{ij}(\omega _{d})\right) \left( \gamma _{cv\mathbf{k}}^{km}(\omega
_{d}^{\prime })\right) ^{\ast }\delta \left( 2\omega _{a}-\omega _{cv}(%
\mathbf{k})\right) \right) \\
&&\times E^{i}(\omega _{a}-\frac{1}{2}\omega _{d})E^{j}(\omega _{a}+\frac{1}{%
2}\omega _{d})\;\left( E^{l}(\omega _{a}-\frac{1}{2}\omega _{d}^{\prime
})E^{m}(\omega _{a}+\frac{1}{2}\omega _{d}^{\prime })\right) ^{\ast }.
\end{eqnarray*}%
Clearly the $\omega _{a}$ integral can be restricted to positive
frequencies, and we can write 
\begin{eqnarray}
\Delta n  \label{2pnoDC} 
&=&\int_{0}^{\infty }d\omega _{a}\int d\omega _{d}d\omega _{d}^{\prime
}\left( \sum_{c,v}\int \frac{d\mathbf{k}}{8\pi ^{3}}\frac{1}{2}\left( \gamma
_{cv\mathbf{k}}^{ij}(\omega _{d})\right) \left( \gamma _{cv\mathbf{k}%
}^{lm}(\omega _{d}^{\prime })\right) ^{\ast }\delta \left( 2\omega
_{a}-\omega _{cv}(\mathbf{k})\right) \right)  \nonumber \\
&&\times E^{i}(\omega _{a}-\frac{1}{2}\omega _{d})E^{j}(\omega _{a}+\frac{1}{%
2}\omega _{d})\;\left( E^{l}(\omega _{a}-\frac{1}{2}\omega _{d}^{\prime
})E^{m}(\omega _{a}+\frac{1}{2}\omega _{d}^{\prime })\right) ^{\ast }, 
\nonumber
\end{eqnarray}
which is the zero DC field equivalent of Eq.~(12).

Using Eq.~(21) we find
\begin{equation}
\frac{dn}{dt}=16\pi ^{3}\left( \sum_{c,v}\int \frac{d\mathbf{k}}{8\pi ^{3}}%
\frac{1}{2}\left( \gamma _{cv\mathbf{k}}^{ij}(0)\right) \left( \gamma _{cv%
\mathbf{k}}^{lm}(0)\right) ^{\ast }\delta \left( 2\omega _{o}-\omega _{cv}(%
\mathbf{k})\right) \right) E_{o}^{i}E_{o}^{j}\left(
E_{o}^{l}E_{o}^{m}\right) ^{\ast }.  \label{dndt2noDC}
\end{equation}

\section{Bichromatic coherent control with no DC field}

To work out the interference term in the carrier injection rate with no DC field, we combine the first-order wavefunction\cite{wahlstrand_independent-particle_2010}
\begin{equation}
\left\vert \Psi ^{(1)}\right\rangle =\sum_{c,v,\mathbf{k}}\int d\omega
\;\gamma _{cv\mathbf{k}}^{i}E^{i}(\omega )\delta (\omega -\omega _{cv}(%
\mathbf{k}))\left\vert cv(\mathbf{k})\right\rangle ,
\label{psi1noDC}
\end{equation}
and Eq.~(\ref{psi2noDC})
to give
\begin{eqnarray*}
\left\langle \Psi^{(2)}|\Psi^{(1)}\right\rangle 
&=&\sum_{c,v,\mathbf{k}}\int d\omega _{a}d\omega _{d}d\omega \left[ \gamma _{cv%
\mathbf{k}}^{ij}(\omega _{d})\right] ^{\ast }\gamma _{cv\mathbf{k}}^{l} \\
&&\times \left[ E^{i}(\omega _{a}-\frac{1}{2}\omega _{d})E^{j}(\omega _{a}+%
\frac{1}{2}\omega _{d})\right] ^{\ast }E^{l}(\omega )\delta (2\omega
_{a}-\omega _{cv}(\mathbf{k}))\delta (\omega -\omega _{cv}(\mathbf{k})) \\
&=&\sum_{c,v,\mathbf{k}}\int d\omega _{a}d\omega _{d}\left[ \gamma _{cv\mathbf{%
k}}^{ij}(\omega _{d})\right] ^{\ast }\gamma _{cv\mathbf{k}}^{l}\delta
(2\omega _{a}-\omega _{cv}(\mathbf{k})) \\
&&\times \left[ E^{i}(\omega _{a}-\frac{1}{2}\omega _{d})E^{j}(\omega _{a}+%
\frac{1}{2}\omega _{d})\right] ^{\ast }E^{l}(2\omega _{a}),
\end{eqnarray*}
so 
\begin{multline}
\Delta N_{(I)} =\sum_{c,v,\mathbf{k}}\int d\omega _{a}d\omega _{d}\left[ \gamma _{cv\mathbf{%
k}}^{ij}(\omega _{d})\right] ^{\ast }\gamma _{cv\mathbf{k}}^{l}\delta
(2\omega _{a}-\omega _{cv}(\mathbf{k})) \\
\times \left[ E^{i}(\omega _{a}-\frac{1}{2}\omega _{d})E^{j}(\omega _{a}+%
\frac{1}{2}\omega _{d})\right] ^{\ast }E^{l}(2\omega _{a}) +c.c., \nonumber
\end{multline}%
and, converting the sum over $\mathbf{k}$ to an integral,%
\begin{multline}
\Delta n_{(I)}  \label{deltaninter} =\int d\omega _{a}d\omega _{d}\left( \sum_{c,v}\int \frac{d\mathbf{k}}{8\pi
^{3}}\left[ \gamma _{cv\mathbf{k}}^{ij}(\omega _{d})\right] ^{\ast }\gamma
_{cv\mathbf{k}}^{l}\delta (2\omega _{a}-\omega _{cv}(\mathbf{k}))\right)  \\
\times \left[ E^{i}(\omega _{a}-\frac{1}{2}\omega _{d})E^{j}(\omega _{a}+%
\frac{1}{2}\omega _{d})\right] ^{\ast }E^{l}(2\omega _{a})
+c.c.,
\end{multline}
which is the zero field equivalent of Eq.~(15).

Here we consider the optical field given by Eq.~(28), and we consider carrier
injection due to one- and two-photon absorption, and the interference
between those processes, 
\[
\frac{dn}{dt}=\frac{dn_{(1)}}{dt}+\frac{dn_{(2)}}{dt}+\frac{dn_{(I)}}{dt}. 
\]
Here we are only concerned with the interference term $dn_{(I)}/dt$.
We use Eq.~(30) in (\ref{deltaninter}) to
give 
\begin{equation}
\frac{dn_{(I)}}{dt}
=4\pi ^{2}\left( \sum_{c,v}\int \frac{d\mathbf{k}}{8\pi ^{3}}\left[ \gamma
_{cv\mathbf{k}}^{ij}(0)\right] ^{\ast }\gamma _{cv\mathbf{k}}^{l}\delta
(2\omega _{o}-\omega _{cv}(\mathbf{k}))\right) \left[ E_{F}^{i}E_{F}^{j}%
\right] ^{\ast }E_{S}^{l} +c.c.
\end{equation}